\documentclass{jfm}
\usepackage{comment}
\usepackage{graphicx}
\usepackage{newtxtext}
\usepackage{newtxmath}
\usepackage{natbib}
\usepackage{hyperref}
\usepackage{subcaption}
\usepackage{xcolor}
\usepackage{mathMacros}
\usepackage{siunitx}
\usepackage{mathtools}
\usepackage{amsmath}
\usepackage{natbib}
\hypersetup{
    colorlinks = true,
    urlcolor   = blue,
    citecolor  = black,
}

\newcommand{\RomanNumeralCaps}[1]
\linenumbers

\definecolor{minty}{RGB}{170, 255, 195}
\definecolor{mauve}{RGB}{224, 176, 255}
\definecolor{cinnamon}{RGB}{210, 105, 30}
\usepackage[most]{tcolorbox}
\newtcolorbox{tiBox}{
  colback=minty,   
  colframe=black,      
  coltext=black,       
  boxrule=0.8pt,       
  arc=1mm,             
  left=2pt, right=2pt, top=2pt, bottom=2pt, 
  boxsep=1pt,          
  enhanced,            
  sharp corners        
}

\title{Solid adsorption: the missing mechanism for surfactant contact lines - a phase-field approach }

\author{Parvathy K. Kannan\aff{1}
  \corresp{\email{pakk@kth.se}},
  Kazi T. Iqbal\aff{1}, Diego D\'iaz\aff{1}, Ilse Mateman\aff{2}, Shahab Mirjalili\aff{1}, Gustav Amberg\aff{1}, Shervin Bagheri\aff{1}
 \and Outi Tammisola\aff{1}}

\affiliation{\aff{1}FLOW, Department of Engineering Mechanics, KTH Royal Institute of Technology, SE-100 44 Stockholm, Sweden
\aff{2} Department of Mechanical Engineering, Eindhoven University of Technology, Netherlands
}

\begin{document}
\maketitle

\begin{abstract}
We develop a thermodynamically consistent phase-field model for soluble surfactants in two-phase flows, incorporating both interfacial and solid surface adsorption. The model is derived via variational principles consistent with the second law of thermodynamics, resulting in modified free energies and boundary conditions that capture surfactant transport, adsorption, and wetting dynamics. A key contribution of this work is the inclusion of surfactant adsorption on solid walls, which leads to qualitative agreement with experimental observations: unlike prior numerical studies that predicted hydrophilic surfaces becoming more hydrophilic and hydrophobic surfaces more hydrophobic, our model shows a shift toward increased hydrophilicity across all contact angles-consistent with experimental trends. Our results establish that solid adsorption provides the missing mechanism required for predictive modelling of surfactant-laden contact line dynamics.
\end{abstract}

\begin{keywords}
Authors should not enter keywords on the manuscript, as these must be chosen by the author during the online submission process and will then be added during the typesetting process (see \href{https://www.cambridge.org/core/journals/journal-of-fluid-mechanics/information/list-of-keywords}{Keyword PDF} for the full list).  Other classifications will be added at the same time.
\end{keywords}

{\bf MSC Codes }  {\it(Optional)} Please enter your MSC Codes here

\section{Introduction}
\label{sec:intro}

Surfactants play a pivotal role in multiphase systems by lowering interfacial tension and thereby modifying wetting and spreading behaviour. This ability underpins a wide range of applications, from enhanced oil recovery \citep{osama2020,liang2023} and coating flows to biomedical processes \citep{ceresa2021}. Soluble surfactants are of particular interest since they partition between the bulk and the interface, dynamically altering interfacial properties and driving complex interfacial flows \citep{afsar2003unstable}.

\begin{figure}
  \centerline{\includegraphics[width=0.7\textwidth]{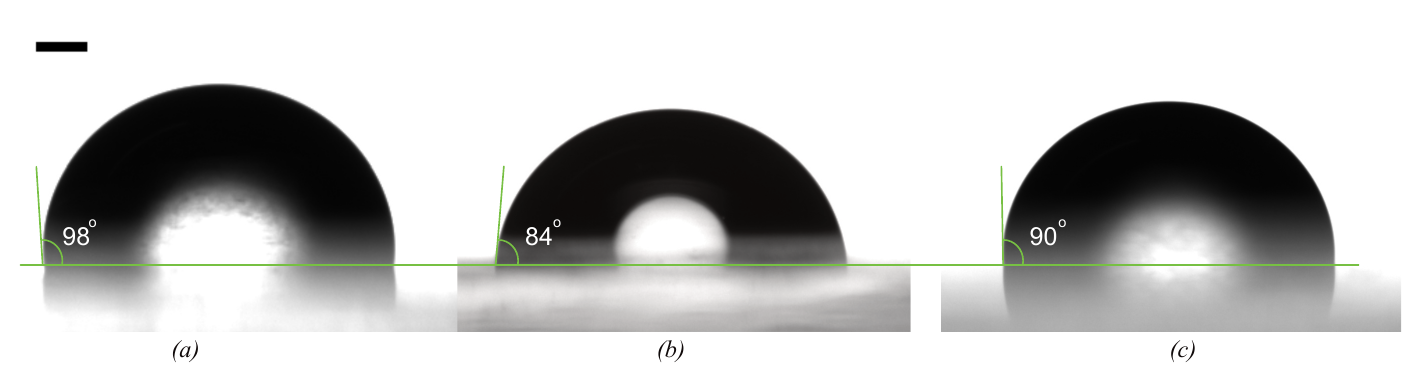}}
  \caption{Equilibrium contact angle of a spreading droplet. (a) water (b) 0.1 $\%$ PEO solution (c) 0.03 $\%$ PEG solution on a PDMS surface. The scale bar represents 0.5 mm. }
\label{fig:exp-st}
\end{figure}

A robust experimental finding-observed across diverse surfactant-substrate systems is that surfactants systematically reduce equilibrium contact angles: both hydrophilic and hydrophobic surfaces shift toward increased wettability \citep{stoebe1996surfactant, lee2008kinetics}. Figure~\ref{fig:exp-st} illustrates this universal trend using our own experimental measurements of equilibrium contact angles for water, 0.1\% PEO solution, and 0.03\% PEG solution on a PDMS substrate, where we observe a clear reduction in contact angle upon surfactant addition. Even though they are not classical surfactants with hydrophobic tails, they are surface active and reduce the interfacial tension by adsorbing on the interface. Experimental results of spreading of droplets upon addition of classical surfactants line Triton X-100 and Poloxamer are shown in Appendix \ref{exp-res}. Beyond this universal trend, surfactants can also produce counterintuitive behaviours such as autophobing, where droplets spontaneously recoil upon surfactant addition \citep{bera2016surfactant, tadmor2019drops}. These phenomena point to a mechanism beyond simple interfacial tension reduction: wall adsorption must play a central role.

Over the last two decades, several numerical frameworks have been developed to model soluble surfactants in multiphase flows. \citet{xu2018} introduced a level-set formulation for soluble surfactants, \citet{khatri2014} developed an interface-tracking approach, while diffuse-interface methods were pursued by \citet{teigen2014, Engblom2013}. Building on these, \citet{zhu2019surfmain} extended a phase-field method to include soluble surfactants in moving contact line problems. Phase-field models, particularly those based on the Cahn--Hilliard formulation, have become increasingly popular due to their thermodynamic consistency and their ability to regularize the singular stresses at the contact line \citep{Jacqmin2000, Yue2020cah, qiu2025phase}. Yet despite their sophistication, all existing formulations on solid substrates share a common limitation.

All formulations on solid substrates neglect surfactant adsorption at walls. Without wall adsorption, surfactants only reduce fluid--fluid interfacial tension $\sigma$, rescaling Young's equation as $\sigma \cos \theta = \sigma_0 \cos \theta_0$, where $\theta_0$ is the equilibrium angle for the clean system and $\theta$ is the equilibirum angle of the contaminated system. This rescaling amplifies the substrate's wetting character---hydrophilic substrates become more hydrophilic, hydrophobic substrates more hydrophobic---but cannot produce the universal shift toward hydrophilicity observed experimentally. Nor can it explain autophobing, which requires asymmetric modification of the solid–liquid and solid–gas interfacial tensions. This fundamental mismatch between model predictions and experimental observations reveals that a critical mechanism is missing.

In this work, we provide the missing mechanism i.e. solid adsorption by developing a thermodynamically consistent phase-field model that incorporates surfactant adsorption on solid walls in addition to interfacial adsorption. We demonstrate that this single addition—solid surface adsorption—resolves the qualitative discrepancy between prior numerical predictions and experiments: our model predicts a universal shift toward hydrophilicity across all contact angles \citep{staniscia2022tuning}, consistent with experimental trends.

We validate the model against clean-droplet spreading benchmarks, derive scaling laws for the solid adsorption layer, and compare predictions with experimental measurements of polymeric surfactants, including autophobing phenomena. We further illustrate the role of wall adsorption in Couette flow, where we find that only preferential adsorption on the solid–liquid interface enhances droplet stability under shear, extending the critical capillary number by 20–30\%.

This paper is organized as follows. In \S\ref{sec:free-en} we introduce the phase-field framework: the bulk free energy and the resulting Cahn–Hilliard–Navier–Stokes equations (\S\ref{sub:gov}). The wall free energy is formulated in \S\ref{sec:PF wall}, resulting in kinetic boundary conditions for surfactant transport and contact line movement. Equilibrium properties are analysed at the fluid–fluid interface (\S\ref{sec:eq-pro}) and for the solid-adsorption layer (\S\ref{surf-bl}). The numerical scheme is described in \S\ref{sec:num} and the experimental methods are given in Appendix \ref{sec:exp-methods}. Results are presented in \S\ref{sec:results}, progressing from validation through spreading dynamics, scaling law verification, and Couette flow stability. Conclusions are summarized in \S\ref{sec:con}.

\section{Phase-field method}
\label{sec:free-en}

\begin{figure}
  \centerline{\includegraphics[width=0.6\textwidth]{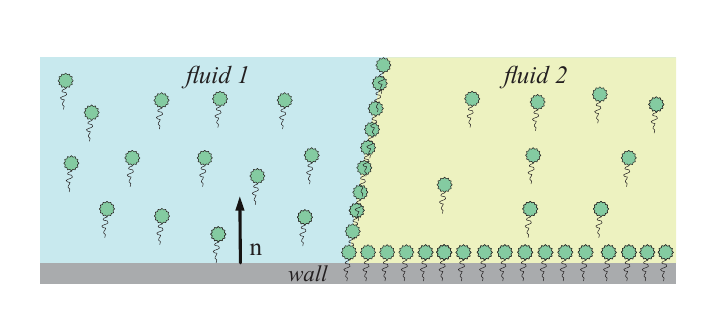}}
  \caption{Illustration of a surfactant-laden fluid-fluid system on a solid substrate }
\label{fig:phasefield}
\end{figure}
We consider an incompressible two-phase system with $\phi$ being the phase field variable ranging from -1 in fluid 1 to +1 in fluid 2 via a diffuse interface of thickness $\epsilon$ as shown in Figure \ref{fig:phasefield}. The free energy is a combination of three energies excluding the kinetic energy, namely the mixing energy, the energy due to the presence of surfactants, and the surface energy. We will discuss the surface energy extensively in \S \ref{sec:PF wall}.

\subsection{Free energy of the bulk}
\label{sec:freeenergy}
The mixing energy per unit volume \citep{CahnHilliard1958} can be written as
\begin{equation}
  F_m(\phi, \nabla \phi)=\lambda \left( \frac{1}{2}|\nabla \phi|^2 + F_0(\phi)\right) \;,
  \label{fm}
\end{equation}
 where

\begin{equation}
  F_0(\phi)=\frac{(\phi^2-1)^2}{4\epsilon^2}
  \label{f0}
\end{equation}
 is the double-well potential with minima in the bulk phases (-1 and +1) thereby preferring the separation of the system into two phases. The square gradient term of the mixing energy $\lambda |\nabla \phi|^2 /2 $ on the other hand prefers complete mixing of the two phases. Here $\lambda$ is the mixing energy density often defined as

 \begin{equation}
  \lambda= \frac{3}{2\sqrt{2}}\sigma \epsilon \;.
  \label{lambda}
\end{equation}
where $\sigma$ is the interfacial tension between the two fluids.

In the presence of surfactants, additional energetic contributions arise from spatial inhomogeneity and bulk mixing entropy. Following \citet{Engblom2013}, we adopt a logarithmic free-energy density per unit volume,
\begin{equation}
  F_{\psi}(\psi,\nabla\psi)
  = \frac{\lambda}{\epsilon^{2}}\,\text{Pi}\,
    \big[\,\psi\ln\psi + (1-\psi)\ln(1-\psi)\,\big]
  \;+\; \frac{\lambda_{s}}{2}\,|\nabla \psi|^{2},
  \label{eq:Fpsi}
\end{equation}
where $\psi\in[0,1]$ is the local surfactant concentration, Pi is a temperature-dependent energetic scale tied to the diffusive entropy of the mixture \citep{Engblom2013}, and $\lambda_s>0$ is a gradient-penalty coefficient.

The logarithmic term encodes the \emph{entropy of mixing} of surfactant and solvent: it is strictly convex on $(0,1)$, discourages unphysical saturation ($\psi\to 0$ or $1$), and guarantees a thermodynamically consistent chemical potential. The gradient term $\lambda_s|\nabla\psi|^{2}$ regularizes sharp concentration fronts by penalizing steep gradients; physically it sets a finite adsorption-layer thickness and mathematically it ensures well-posedness of the coupled transport by contributing a $-\lambda_s\nabla^2\psi$ term to the chemical potential. We return to the far-from-wall equilibrium profile implied by \eqref{eq:Fpsi} in \S\ref{sec:eq-pro}. Critically, its presence also provides the mathematical structure necessary for the surface adsorption boundary condition derived in \S\ref{sec:chnsbc}, representing a key difference from prior work \citep{zhu2019surfmain}. 

The decrease of free energy per unit volume due to adsorption of surfactants onto the interface is governed by $F_1$ given as
\begin{equation}
  F_{1}(\phi,\psi)= -\frac{\lambda}{4\epsilon^2} \psi (1-\phi^2)^2\;.
  \label{F1}
\end{equation}
Finally, complementary to $F_1$, the free surfactants in the bulk are penalised by an increase in free energy per unit volume, $F_{\text{Ex}}$, given as
\begin{equation}
  F_\text{Ex}(\phi,\psi)= \frac{\lambda}{2 \text{Ex}\epsilon^2} \psi \phi^2 \;.
  \label{Fex}
\end{equation}

where $\text{Ex}$ is the constant related to how soluble the phases are with each other.
Therefore, the total bulk free energy in the presence of surfactants can be given as
\begin{equation}
  F_{bulk}(\phi,\psi,\nabla\phi,\nabla\psi)= F_m(\phi, \nabla \phi)+F_{\psi}(\psi,\nabla\psi)+F_{1}(\phi,\psi)+F_\text{Ex}(\phi,\psi) \;.
  \label{Fbulk}
\end{equation}
\subsection{Cahn-Hilliard Navier-Stokes governing equations}
\label{sub:gov}

From the bulk free energy defined in \S\ref{sec:freeenergy}, we derive transport equations for the phase field variable $\phi$ and surfactant concentration $\psi$ using variational principles. The complete derivation is provided in Appendix~\ref{app:chns_derivation}; here we summarize the results.

\subsubsection{Navier-Stokes Equations}

The incompressible Navier-Stokes equations coupled to the phase-field variables are:
\begin{equation}
\rho\left(\frac{\partial \boldsymbol{u}}{\partial t} + \boldsymbol{u}\cdot\bnabla \boldsymbol{u}\right) + \boldsymbol{J}.\bnabla \boldsymbol{u} = -\bnabla p + \bnabla \cdot \left(\mu\left(\bnabla\boldsymbol{u} + \bnabla\boldsymbol{u}^\mathrm{T}\right)\right) + G_\phi\bnabla \phi + G_\psi\bnabla \psi,
\label{eq:ns_momentum}
\end{equation}
\begin{equation}
\bnabla\cdot\boldsymbol{u} = 0,
\label{eq:ns_continuity}
\end{equation}
where $\rho$ and $\mu$ are density and viscosity interpolated between the two fluids:
\begin{equation}
\rho =\rho_1\frac{ (1+\phi)}{2} + \rho_2\frac{ (1-\phi)}{2},
\label{eq:rho_interp}
\end{equation}
\begin{equation}
\mu =\mu_1\frac{ (1+\phi)}{2} + \mu_2\frac{ (1-\phi)}{2}.
\label{eq:mu_interp}
\end{equation}
$\boldsymbol{J}$ is the diffusive flux caused by deviation of $u$ from mass-averaged velocities \citep{abels2012thermodynamically, zhu2019surfmain} and is expanded to be:
\begin{equation}
\boldsymbol{J} = \frac{\rho_1-\rho_2}{2}M_\phi\bnabla G_\phi,
\label{eq:ns_continuity}
\end{equation}
The terms $G_\phi\bnabla\phi$ and $G_\psi\bnabla\psi$ represent surface tension forces acting at the interfaces and $M_\phi$ is the phase-field mobility.

\subsubsection{Cahn-Hilliard Equation for Phase Field}

The phase field variable $\phi$ (conserved) evolves via:
\begin{equation}
\Ddt{\phi} = M_\phi \bnabla^2 G_\phi,
\label{eq:ch_phi}
\end{equation}
where $G_\phi$ is the chemical potential of the phase field variable:
\begin{equation}
G_\phi =\lambda \left(F_0'(\phi) - \bnabla^2 \phi + \frac{\psi \phi}{\text{Ex}\epsilon^2} + \frac{\psi\phi(1-\phi^2)}{\epsilon^2} \right).
\label{eq:mu_phi}
\end{equation}
Here, $F_0'(\phi) = (\phi^3 - \phi)/\epsilon^2$ is the derivative of the double-well potential. The last two terms arise from variational derivatives of the interfacial adsorption energy $F_1$ and bulk surfactant penalty $F_\text{Ex}$ (see Appendix~\ref{app:chns_derivation}).

\subsubsection{Cahn-Hilliard Equation for Surfactant}

The surfactant concentration $\psi$ (conserved) evolves via:
\begin{equation}
\Ddt{\psi} = \bnabla\cdot (M_\psi \bnabla G_\psi),
\label{eq:ch_psi}
\end{equation}
where $M_\psi = m_\psi \psi(1-\psi)$ is a degenerate mobility that prevents unphysical surfactant concentrations exceeding saturation: the mobility vanishes as $\psi \to 0$ or $\psi \to 1$, enforcing physical bounds on the concentration field. Also, $G_\psi$ is the chemical potential of the surfactant concentration:
\begin{equation}
G_\psi = \frac{\lambda}{\epsilon^2}\left( \text{Pi} \ln \left(\frac{\psi}{1-\psi}\right) + \frac{\phi^2}{2\text{Ex}} - \frac{(1-\phi^2)^2}{4}\right) - \lambda_s \bnabla^2 \psi.
\label{eq:mu_psi}
\end{equation}

The logarithmic term encodes thermodynamic consistency (entropy of mixing), the $\phi$-dependent terms couple surfactant to interfacial position, and the $-\lambda_s\bnabla^2\psi$ term regularizes sharp concentration gradients. This final term is crucial: it introduces a characteristic length scale and necessitates an additional boundary condition at walls (discussed in \S\ref{sec:chnsbc}).

\subsection{Surface energy}
\label{sec:PF wall}

When the diffuse interface intersects a solid surface, contact lines emerge that require special treatment. The thermodynamic treatment of the wall requires a free energy functional that correctly captures the equilibrium contact angle and provides kinetic boundary conditions for dynamic wetting.

For a clean two-phase system without surfactants, the wall free energy per unit area follows the standard form \citep{yue2011wall}:
\begin{equation}
f_w = -\sigma_0 \cos\theta_{S,0} g(\phi) + \frac{\sigma_{w1,0} + \sigma_{w2,0}}{2},
\end{equation}
where $\sigma_{w1,0}$ and $\sigma_{w2,0}$ are the clean surface tensions between the solid and fluids 1 and 2, respectively, and $g(\phi)$ is a switch function that modulates the wall energy smoothly across the diffuse interface. The switch function must satisfy $g(\pm 1) = \pm 0.5$ so that in pure bulk phases, $f_w$ recovers the respective solid-fluid interfacial tensions.

This construction is consistent with Young's equation:
\begin{equation}
\sigma_0 \cos\theta_{S,0} = \sigma_{w1,0} - \sigma_{w2,0}.
\end{equation}

In the presence of surfactants, both fluid-fluid and solid-fluid surface tensions are modified. Therefore, the switch function $g(\phi)$ must be generalized to account for surfactant-induced changes in interfacial energetics.

To incorporate the surfactant effects we follow \cite{Yue2020cah}, and derive $g(\phi)$ to be
\begin{equation}
g(\phi) = \frac{3}{2\sqrt{2}}\sqrt{A} \left[\phi + \frac{B}{6A}\phi^3 + \frac{C}{10A}\phi^5 + \frac{D}{14A}\phi^7\right],
\end{equation}
where
\begin{subequations}
\begin{align}
A &= \frac{1}{2} + \frac{\psi_b}{2} - \frac{\psi_b}{E_x} - \psi_b,\\
B &= -1 + \frac{\psi_b}{E_x} + \psi_b,\\
C &= \frac{1}{2} - \frac{\psi_b}{2},\\
D &= 14A\left[\frac{\sqrt{2}}{3\sqrt{A}} - 1 - \frac{B}{6A} - \frac{C}{10A}\right].
\end{align}
\end{subequations}
The derivation of this modified switch function is given in Appendix \ref{sec:switch}.
When $\psi_b \to 0$ (no surfactant), this reduces to the clean case $g(\phi) = \phi(3-\phi^2)/4$. The surfactant-dependent terms modify the switch function to reflect changes in interfacial energetics induced by adsorption.

\subsubsection{Surface adsorption}  \label{sec:sur-ads}

When surfactants are added to a two-phase system in contact with a solid,they could adsorb on the solid-fluid interfaces, thereby reducing the wall-fluid interfacial tensions \citep{haidara1995direct}.

We model the reduction in solid-fluid surface tension using the Langmuir-Szyszkowski equation of state \citep{soligo2019coalescence}:
\begin{equation}
\sigma_{wf} = \sigma_{wf,0}[1 + \beta\log(1-\psi)],
\label{eq:lang}
\end{equation}
where $\sigma_{wf,0}$ is the clean wall-fluid surface tension, $\beta$ is the elasticity number ($0 < \beta < 1$) controlling surfactant strength, and $\psi$ is the local surfactant concentration at the wall. This logarithmic form, also known as the Szyszkowski equation, is thermodynamically equivalent to the Langmuir adsorption isotherm. The logarithmic dependence emerges from the Gibbs adsorption equation relating surface tension to surface excess concentration.

The Langmuir--Szyszkowski equation is chosen for its thermodynamic consistency and widespread validation for common surfactants at moderate concentrations. This isotherm assumes monolayer adsorption with equivalent, non-interacting adsorption sites. Alternative isotherms can be incorporated into the present framework by replacing equation \ref{eq:lang} with the appropriate equation of state. For dilute surfactant concentrations ($\psi \ll 1$), a linear equation of state $\sigma_{wf} = \sigma_{wf,0}(1 - \beta\psi)$ provides a simpler approximation that is often sufficient. For systems with lateral interactions between adsorbed molecules, the Frumkin isotherm offers a more accurate description, while the Freundlich isotherm is appropriate for heterogeneous surfaces with a distribution of adsorption energies. However, the Langmuir--Szyszkowski form suffices for the phenomena investigated here and we proceed with that.

Incorporating surfactant adsorption on both solid-fluid interfaces into Young's equation gives:
\begin{equation}
\sigma \cos\theta_S = \sigma_0 \cos\theta_{S,0} + \sigma_{w1,0}\beta_{s1}\log(1-\psi) - \sigma_{w2,0}\beta_{s2}\log(1-\psi),
\end{equation}
where $\beta_{s1}$ and $\beta_{s2}$ are the surfactant strengths on the solid-phase 1 and solid-phase 2 interfaces, respectively. Different elasticity numbers allow for asymmetric adsorption behavior on the two interfaces. This asymmetry is crucial for phenomena like autophobing, where differential adsorption on solid-liquid versus solid-gas interfaces can cause droplet recession.

The complete wall energy per unit area becomes:
\begin{equation}   \label{eq:fwfinal}
\begin{aligned} 
   f_w =  -\left[\sigma_0 \cos\theta_{S0} + \log(1-\psi)(\sigma_{w1,0}\beta_{s1}  - \sigma_{w2,0}\beta_{s2})\right] g(\phi) \\ + \log(1-\psi)\frac{\sigma_{w1,0}\beta_{s1} +\sigma_{w2,0}\beta_{s2}}{2} + \frac{\sigma_{w1,0} +\sigma_{w2,0}}{2}\;,
\end{aligned}
\end{equation}

This expression contains three physically distinct contributions: (i) surfactant-modified Young stress modulated by the switch function $g(\phi)$, (ii) average reduction in wall-fluid tensions independent of $\phi$, and (iii) baseline wall energy for the clean system.

Experimental studies indicate that surfactant-induced reduction of wall-fluid surface tension saturates at high coverage and does not decrease indefinitely \citep{chang1995adsorption,ju2017equilibrium,lopez2006surface,soligo2019coalescence}. To prevent unphysical values, we limit the minimum surface tension to half the clean-wall value:
\begin{equation}
\label{eq:cap}
\sigma_{w\alpha}(\psi) = \sigma_{w\alpha,0} \max[0.5, 1 + \beta_{s\alpha}\log(1-\psi)]
\end{equation}
for each wall-fluid interface $\alpha \in \{1,2\}$.

\citet{soligo2019coalescence} note that ``according to experimental observations, surface tension does not decrease below this threshold (roughly).'' The factor 0.5 provides a physically reasonable bound ensuring $\sigma_{w\alpha} \geq 0$ and a smooth transition from the logarithmic regime to a saturated plateau at high $\psi$.

With the wall free energy fully specified, we now derive the corresponding boundary conditions for the phase field and surfactant concentration in \S \ref{sec:chnsbc}.

\subsection{Cahn-Hillard Navier-Stokes boundary conditions}
\label{sec:chnsbc} 
We now collect all the surface integral terms from the variational derivative of the bulk free energy with respect to $\phi$, and add it into the  variational derivative of the surface energy term with respect to $\phi$, to get
\begin{equation}
   \delta  F = \lambda\oint (\nabla\phi.\boldsymbol{n} )\delta \phi))dS  + \oint f'w(\phi)\delta \phi dS
\end{equation}

Requiring that the wall contribution to $\mathrm{d}F/\mathrm{d}t$ be non-positive (thermodynamic dissipation), the relaxation boundary condition for $\phi$ takes the form
\begin{equation}    \label{eq:wet-bc}
    \frac{\partial \phi}{\partial t} + \boldsymbol{u}\cdot\bnabla \phi = -\Gamma L_\phi \; ,
\end{equation}

where $L_{\phi}$ is the wall potential (as introduced earlier) of the phase field variable and is expressed as

\begin{equation}    \label{eq:L_phi}
    L_{\phi}=\lambda \boldsymbol{n} \cdot \bnabla \phi + \frac{\partial f_w(\phi,\psi)}{\partial \phi} \; ,
\end{equation}
where the capping from equation \ref{eq:cap} introduces piecewise behavior in the derivatives $\partial f_w/\partial\phi$. Defining
\begin{equation}
S_1 = 1 + \beta_{s1}\log(1-\psi), \quad S_2 = 1 + \beta_{s2}\log(1-\psi),
\end{equation}
the derivative with respect to $\phi$ becomes (with $K = \sigma_0\cos\theta_{S,0}$):
\begin{equation}
\frac{\partial f_w}{\partial\phi}(\psi,\phi) = g'(\phi) \times 
\begin{cases}
\frac{K}{2}, & \text{if } S_1, S_2 \leq 0.5,\\
K - \sigma_{w1}\beta_{s1}\log(1-\psi) - \frac{\sigma_{w2}}{2}, & \text{if } S_1 \leq 0.5 < S_2,\\
K + \sigma_{w2}\beta_{s2}\log(1-\psi) + \frac{\sigma_{w1}}{2}, & \text{if } S_2 \leq 0.5 < S_1,\\
K + (\sigma_{w2}\beta_{s2} - \sigma_{w1}\beta_{s1})\log(1-\psi), & \text{if } S_1, S_2 > 0.5,
\end{cases}
\end{equation}

and $g'(\phi)$ can be found in equation \ref{eq:gdash}. This governs the behaviour of the contact line on the surface. $\Gamma = 1/\mu_f \epsilon$ in equation \ref{eq:wet-bc} is a positive phenomenological parameter that controls how fast the contact line relaxes to the equilibrium angle $\theta_S$. The variable $\mu_f$ is known as the contact line friction. This boundary condition governs the dynamics of the contact line according to the contact angle.

Similarly, including the surface integral terms from the variational derivative of the free energy with respect to $\psi$, into the  variational derivative of the surface energy term with respect to $\psi$, we get

\begin{equation}
   \delta  F = \lambda_s\oint (\nabla\psi.\boldsymbol{n} )\delta \psi))dS  + \oint f'w(\psi)\delta \psi dS
\end{equation}
Similarly, requiring dissipation of the surface free energy contribution, the relaxation boundary condition for $\psi$ takes the form
\begin{equation}    \label{eq:ads-bc}
    \frac{\partial \psi}{\partial t} + \boldsymbol{u}\cdot\bnabla \psi = -\Gamma_s L_\psi \; ,
\end{equation}

where $\Gamma_s$ is a positive constant determining how fast surfactant adsorbs on the walls and  $L_{\psi}$ is the wall potential of the surfactant concentration variable and is expressed as

\begin{equation}    \label{eq:L_psi}
    L_{\psi}=\lambda_s \boldsymbol{n} \cdot \bnabla \psi + \frac{\partial f_w(\phi,\psi)}{\partial \psi} \; ,
\end{equation}

where the capping introduces piecewise behavior in the derivative $\partial f_w/\partial\psi$. Defining
\begin{equation}
S_1 = 1 + \beta_{s1}\log(1-\psi), \quad S_2 = 1 + \beta_{s2}\log(1-\psi),
\end{equation}

the derivative with respect to $\psi$ is:
\begin{equation}
\frac{\partial f_w}{\partial\psi}(\phi,\psi) = \frac{1}{1-\psi} \times 
\begin{cases}
0, & \text{if } S_1, S_2 \leq 0.5,\\
-\sigma_{w1}\beta_{s1}\left[g(\phi) + \frac{1}{2}\right], & \text{if } S_1 \leq 0.5 < S_2,\\
\sigma_{w2}\beta_{s2}\left[g(\phi) - \frac{1}{2}\right], & \text{if } S_2 \leq 0.5 < S_1,\\
(\sigma_{w2}\beta_{s2} - \sigma_{w1}\beta_{s1})g(\phi) - \frac{\sigma_{w2}\beta_{s2} + \sigma_{w1}\beta_{s1}}{2}, & \text{if } S_1, S_2 > 0.5.
\end{cases}
\end{equation}

When both interfaces saturate ($S_1, S_2 \leq 0.5$), the wall becomes ``surfactant-insensitive'' and $\partial f_w/\partial\psi = 0$. These piecewise expressions recover the uncapped forms when $S_1, S_2 > 0.5$, ensuring thermodynamic consistency while preventing unphysical behavior at high surfactant coverage.

Finally, we have $\boldsymbol{n}\cdot\bnabla \phi = 0 $, $\boldsymbol{n}\cdot\bnabla G_\phi = 0 $, $\boldsymbol{n}\cdot\bnabla G_\psi = 0 $ and $\boldsymbol{u} = \boldsymbol{u}_w$, which are the no flux across walls for $\phi$, no flux of chemical potential across walls for $\phi$ and $\psi$ variables and no slip boundary condition, respectively.

The gradient penalty term $\frac{\lambda_s}{2}|\nabla\psi|^2$ introduced 
in $\S$\ref{sec:freeenergy} is structurally critical to the model, not merely a 
regularisation term. Its variational derivative contributes a $-\lambda_s\nabla^2\psi$ 
term to the surfactant chemical potential $G_\psi$ (equation~\ref{eq:mu_psi}), which is 
the only term in $G_\psi$ that carries spatial information about $\psi$ itself; without 
it, $G_\psi$ reduces to a purely local algebraic function of $\psi$ and $\phi$, with no 
coupling between neighbouring values of the concentration field. More importantly, the 
surface integral arising from the variational derivative of $F_\psi$ with respect to 
$\psi$ produces the $\lambda_s\mathbf{n}\cdot\nabla\psi$ term in the wall potential 
$L_\psi$ (equation~\ref{eq:L_psi}), which is the mechanism by which the wall affinity 
$\beta_s$ prescribes a non-trivial normal gradient of $\psi$ at the boundary. In the 
absence of the $|\nabla\psi|^2$ term, both contributions vanish simultaneously: the 
boundary condition collapses to the purely algebraic condition $\partial f_w/\partial\psi 
= 0$, which imposes no flux, and the Cahn--Hilliard equation for $\psi$ becomes 
second-order in space, requiring only the no-flux condition on $G_\psi$ already satisfied 
at the wall. The wall energy can influence the contact line through $f_w$, but has no 
mathematical pathway to alter the bulk surfactant distribution. The $|\nabla\psi|^2$ term 
is therefore what promotes the surfactant equation to fourth order in space and endows the 
wall boundary condition with the structure necessary to communicate solid adsorption into 
the bulk --- a key distinction from models that omit it \citep{zhu2019surfmain}.

\subsection{Equilibrium properties of the surfactant adsorption on the fluid-fluid interface}
\label{sec:eq-pro}

Since surfactant gets adsorbed onto the fluid-fluid interface, this presents the opportunity of finding the equilibrium properties for $\phi$ and $\psi$, and see if there is a valid adsoprtion isotherm (\cite{Engblom2013}, \cite{zhu2019surfmain}). As stated previously, since the chemical potential of the phase field variable, $G_\phi$ remains unchanged from \cite{zhu2019surfmain}, at steady state, the equilibirum profile of $\phi$ is 
\begin{equation}    \label{eq:steady-state-phi}
    \phi(x) = \phi_b \tanh\left({\phi_b\sqrt{1-\psi_b}\frac{x}{\sqrt{2}\epsilon}}\right)\; ,
\end{equation}

\begin{equation}    \label{eq:phib}
    \phi_b^2 = 1+\frac{1}{\text{Ex}} - \frac{1}{\text{Ex}(1-\psi_b)}\; ,
\end{equation}

where $\phi_b$ is the phase field variable in the bulk.

Now let us derive the equilibirum profile for $\psi$. The chemical potential $G_\psi$ at any location can be written as

\begin{equation}    \label{eq:chemical-potential-surf-eq}
    G_\psi = \frac{\lambda}{\epsilon^2}\left( \text{Pi} \ln \left(\frac{\psi}{1-\psi}\right) + \frac{\phi^2}{2\text{Ex}} + \frac{(1-\phi^2)^2}{4}\right) - \lambda_s \bnabla^2 \psi \; ,
\end{equation}

and in the bulk, $G_\psi$ is
\begin{equation}    \label{eq:chemical-potential-surf-eq-bulk}
    G_{\psi_b} = \frac{\lambda}{\epsilon^2}\left( \text{Pi} \ln \left(\frac{\psi_b}{1-\psi_b}\right) + \frac{\phi_b^2}{2\text{Ex}} + \frac{(1-\phi_b^2)^2}{4}\right)\; ,
\end{equation}

where, $\bnabla^2 \psi$ vanishes in the bulk away from the wall.
 Subtracting equation \ref{eq:chemical-potential-surf-eq} and equation \ref{eq:chemical-potential-surf-eq-bulk}, and introducing an intermediate variable $\psi_c$, we get
\begin{equation}    \label{eq:psic}
    \text{Pi}\log \psi_c(x) =   \frac{\lambda}{\epsilon^2}\left(\frac{\phi_b^2-\phi^2}{2\text{Ex}} + \frac{1}{4}(\phi^2-\phi_b^2)(\phi_b^2+\phi^2-1) \right)- \lambda_s \bnabla^2 \psi \; ,
\end{equation}

The steady state profile is given as

\begin{equation}
    \psi(x) = \frac{\psi_b}{\psi_b+\psi_c(x)(1-\psi_b)}=\frac{\psi_b}{\psi_b+\psi_c(x)} + \mathcal{O}(\psi_b) 
\end{equation}
 As $x\rightarrow0$ along the interface, we have $\phi=0$, phase bulk value $\phi_b=\pm1$, and $\psi_b\ll1$. Therefore,
\begin{equation}    \label{eq:psi_eq0}
    \psi_0 = \frac{\psi_b}{\psi_cR+\psi_b} + \mathcal{O}({\psi_b}) \; ,   
\end{equation}
where $\psi_c$ is the Langmuir adsorption constant and is defined as 
\begin{equation} \label{eqbm-zhu}
    \text{Pi}\log \psi_c = -\frac{1}{4}\left(1+ \frac{2}{\text{Ex}}\right)
\end{equation}
 and $R=\exp(-\lambda_s\epsilon^2\bnabla^2\psi_0/\text{Pi} \lambda )$. When $R=1$, we get the Langmuir isotherm. We are unable to obtain a explicit adsorption isotherm since we do not know the relation between $\psi_0$ and $\bnabla^2\psi_0$. We shall explore numerically further in detail in Appendix \ref{ap:flfl} and see that it does not deviate much from the standard Langmuir isotherm shown by \cite{Engblom2013}.

\subsection{Equilibrium properties of the surfactant adsorption on the solid-fluid interface}
\label{surf-bl}

\begin{figure}
  \centerline{\includegraphics[width=0.6\textwidth]{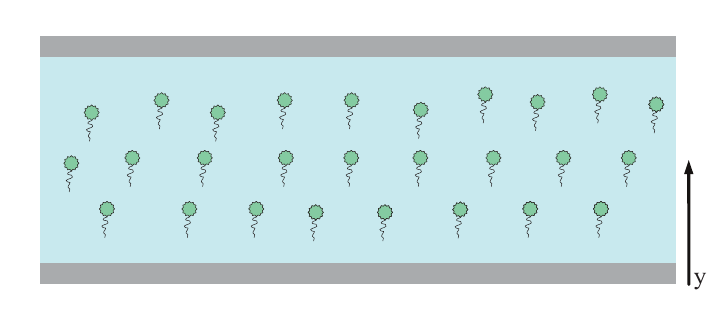}}
  \caption{Illustration of the initial state of a surfactant-laden single-phase system with walls on both ends. }
\label{fig:singlephase}
\end{figure}

In order to characterize the behaviour of the solid adsorption layer, we consider a single phase system with walls (as shown in figure \ref{fig:singlephase} ) at steady state. The Cahn-Hilliard equation at equilibrium can thus be written as,

\begin{equation}    \label{eq:chem_pot0}
    \bnabla \left(\psi(1-\psi)\bnabla \left ( \frac{\lambda}{\epsilon^2}\left( \text{Pi} \ln \left(\frac{\psi}{1-\psi}\right) + \frac{\phi^2}{2Ex} + \frac{(1-\phi^2)^2}{4}\right) - \lambda_s \bnabla^2 \psi \right)\right) = 0, \;
\end{equation}

 In the bulk, we neglect the terms acting only on the interface, and simply consider the terms which dominate the surfactant behavior near the wall. Therefore we can reduce the equation \ref{eq:chem_pot0} to one dimension in the direction of wall to
\begin{equation}    \label{eq:chem_pot1}
    \od{}{y} \left(  \frac{\lambda}{\epsilon^2}\text{Pi} \psi_y - \psi(1-\psi)\lambda_s  \psi_{yyy} \right) = 0, \;
\end{equation}

At $y=\infty$, $\psi=\psi_b$ and $d\psi/dy=\psi_y=0$

Assuming that $\psi_b \ll1$, $\psi_y \approx \Delta\psi/\delta$, where $\Delta \psi$ is the variation in $\psi$ over a thickness $\delta$, and $\psi=\Delta \psi + \psi_b$, we get the following relation at $y=0$:

\begin{equation}    \label{eq:scaling1}
  \frac{\lambda \text{Pi} }{\epsilon^2} \sim \frac{\lambda_s(\Delta \psi + \psi_b)}{\delta^2} \;
\end{equation}

Also from equation \ref{eq:L_psi} at steady state, we know that at $y=0$
\begin{equation}    \label{eq:scaling2}
 \psi_y = \frac{\Delta\psi}{\delta} \sim \frac{\sigma_s\beta}{\lambda_s} \;
\end{equation}

where $\sigma_s$ and $\beta_s$ is the solid-fluid interfacial tension and surfactant strength respectively.
We consider two regimes.

Regime 1: $\Delta\psi \ll \psi_b$

In the case of a small gradient of $\psi$ near the wall, we can approximate $\Delta \psi + \psi_b \approx \psi_b$. We get from equation \ref{eq:scaling1} and \ref{eq:scaling2}
\begin{equation}    \label{eq:scaling4}
\delta \sim \epsilon\sqrt\frac{\psi_b \lambda_s}{\lambda \text{Pi}}
\end{equation}
We notice that when the gradient of surfactant concentration at the wall is small, the thickness of the surfactant layer depends on the surfactant diffusion term, $\lambda_s$ and not the wall surfactant strength, $\beta$. 

Regime 2: $\Delta\psi \gg \psi_b$

When there is a large gradient of $\psi$ near the wall, we can approximate $\Delta \psi + \psi_b \approx \Delta \psi$. From equation \ref{eq:scaling1} and \ref{eq:scaling2}, we get  
\begin{equation}    \label{eq:scaling3}
\delta \sim \frac{\epsilon^2\sigma_s \beta}{\lambda \text{Pi}}
\end{equation}
Here we see that the surfactant layer thickness at the wall depends on the surfactant strength at the wall, $\beta$.

The surfactant gradient at the wall, $\Delta \psi$ can be solved to be
 \begin{equation} \label{eq:scaling2.5}
 \Delta \psi \sim \frac{\epsilon^2 \sigma_s^2 \beta^2}{\text{Pi} \lambda\lambda_s}
 \end{equation}
We shall verify the scalings \ref{eq:scaling4}, \ref{eq:scaling3} and \ref{eq:scaling2.5} numerically in $\S$ \ref{sec:num-res}.

\section{Numerical Scheme}
\label{sec:num}

\subsection{Spatial Discretization}

The coupled Cahn–Hilliard, Navier–Stokes, and surfactant transport equations are solved on a staggered Cartesian grid, with velocity components at cell faces and scalar quantities (pressure, phase field, chemical potential, surfactant concentration) at cell centers. Spatial derivatives are approximated using second-order central finite differences; the domain is periodic in the horizontal direction with the aforementioned boundary conditions applied at the top and bottom walls for the phase-field and surfactant concentration.

\subsubsection{Variable Coefficient Discretization}

To maintain stability when discretizing the Cahn–Hilliard equation for the surfactant field $\psi$, which involves a variable-coefficient Laplacian due to the degenerate mobility $M_\psi$, we employ a directionally dependent forward-backward splitting strategy \citep{leveque2007finite}. This approach isolates the variable coefficient multiplication and respects the wall orientation, preventing spurious oscillations near domain boundaries.

For the upper half of the domain, where the wall normal points upward:
\begin{align*}
\bnabla\cdot M_\psi \bnabla G_\psi \approx D_- M_\psi D_{+} G_\psi
\end{align*}

For the lower half of the domain, where the wall normal points downward:
\begin{align*}
\bnabla\cdot M_\psi \bnabla G_\psi \approx D_+ M_\psi D_{-} G_\psi
\end{align*}

Here, where $D_-$ and $D_+$ are the backwards and forwards difference operators, and $M_\psi$ is the mobility coefficient varying with respect to $\psi$. The ordering reversal at opposite walls ensures that the differencing stencil respects the direction of information flow dictated by wall orientation.


\subsection{Temporal Integration and Solver}

All terms are advanced in time using a second-order Adams–Bashforth scheme. The incompressible Navier–Stokes equations are integrated via a fractional-step method, with the pressure Poisson equation solved using the FFT- based fast Poisson solver \citep{costa2018fft,dodd2014fast}, ensuring an exactly divergence-free velocity field. The clean case without surfactants has been validated and tested in \citet{shahmardi2021fully}. We present them briefly in \S\ref{sec:num-res} to set the problem statement of our surfactant laden model.

\begin{figure}
    \centering
    \includegraphics[width=0.55\textwidth]{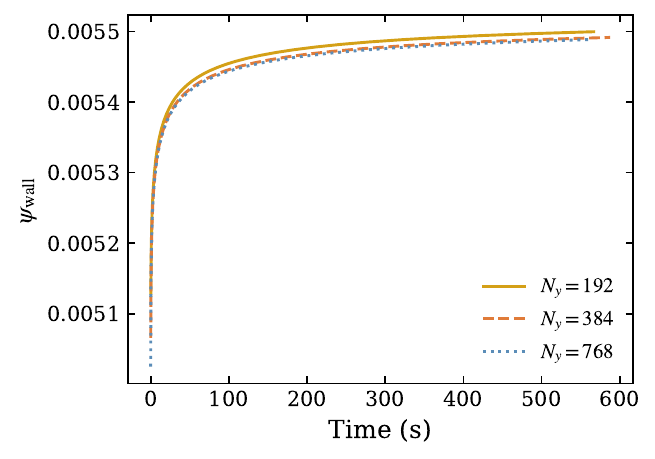}
    \caption{Time evolution of the wall surfactant concentration $\psi_{\mathrm{wall}}$ for three grid resolutions in a 1D single-phase simulation with walls on both ends. The solutions at $N_y = 384$ and $N_y = 768$ are in excellent agreement throughout, confirming grid convergence at the resolution used in the production simulations.}
    \label{fig:grid_convergence}
\end{figure}

The simulations presented in \S\ref{sec:results} use a grid resolution of $N_x \times N_y = 768 \times 768$ for the droplet spreading cases and $N_x \times N_y = 192 \times 384$ for the Couette flow cases. Grid convergence was verified using a 1D single-phase simulation with walls on both ends, comparing the time evolution of $\psi_{\mathrm{wall}}$ at three resolutions ($N_y = 192$, $384$, and $768$), as shown in Figure~\ref{fig:grid_convergence}. The solutions at $N_y = 384$ and $N_y = 768$ are in excellent agreement throughout, while $N_y = 192$ shows a small deviation at late times only. The grid spacing $\Delta y$ provides the physically meaningful link to the 2D production simulations: for the droplet spreading cases the domain height matches the 1D study and $N_y = 768$ corresponds to the finest converged resolution; for the Couette flow cases the domain height is $1/10$ of the 1D domain with $N_y = 192$, yielding a $\Delta y$ approximately five times finer than the converged $N_y = 384$ spacing. In both cases the production simulations are at least as well resolved as the converged grid identified here.
A formal sharp-interface limit for the wall adsorption boundary condition, including the correct joint scaling of $M_\psi$ and $\lambda_s$ with $\mathrm{Cn}$ — analogous to the mobility--Cahn number scaling required for the phase field equation — has not been established for diffuse-interface surfactant models and remains a problem for future work.

This paper will explore two dimensional problems. Extension to three dimensions and incorporation of other effects like contact angle hysteresis are straightforward within the present formulation and will be addressed in forthcoming publications. Also,  we assume matched density and viscosity between the two phases, which simplifies the numerical treatment but excludes systems with large property contrasts such as water–air.

\section{Results and discussion}
\label{sec:results}

We combine thermodynamically consistent phase-field simulations of soluble surfactants with wall adsorption (\S\ref{sec:free-en}–\ref{surf-bl}) and static contact-angle measurements on surfactant solutions (Appendix \ref{sec:exp-methods}) to qualitatively test the model against experiments.  

In this section we present a systematic progression of results. We begin by validating our model against analytical benchmarks for clean droplet spreading. Next, we examine soluble surfactant effects without wall adsorption, reproducing trends previously reported in the literature. Building on this, we incorporate solid adsorption and reveal new interfacial features such as curvature near the contact line and modified equilibrium angles. We then quantify these effects via scaling laws and qualitative comparison with experiments. Finally, we extend the analysis to droplets in external shear, highlighting the interplay between wall adsorption and hydrodynamic Marangoni stresses.

\subsection{Numerical Validation}
\label{sec:num-res}
\begin{figure}
  \centerline{\includegraphics[width=1\textwidth]{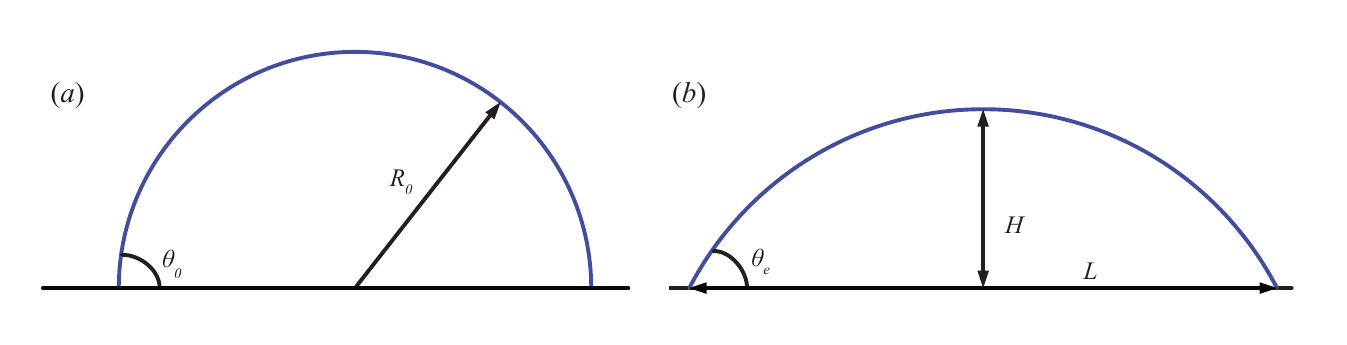}}
  \caption{Schematic of a spreading droplet. a) Initial state of the droplet with radius $R_0$ and initial angle $\theta_0$. b) Equilibrated droplet with contact angle $\theta_e$, contact length $L$, and droplet height $H$. }
\label{fig:sl}
\end{figure}
Following the approach taken by \cite{zhu2019surfmain}, we will first validate the moving contact line of a clean droplet with the analytical solution.  A droplet of radius $R_0=0.35H$ is initialized at $\theta_0=90^\circ$ in a domain of size of $10H$ x $3H$. Gravitational effects are neglected. The Cahn number $\text{Cn} =0.02$ represents the interfacial thickness. The Reynolds number Re$=\rho U_{ref}R_0/\mu$, the capillary number Ca$=\mu U_{ref}/\sigma_0$, and the Peclet number Pe$=U_{ref}R_0\epsilon^2/M_\phi\lambda$ are $10$, $0.1$ and $10^4$, respectively, as defined in \cite{zhu2019surfmain}. The fluids are matched in density and viscosity. The surfactant is initialized according to the equilibrium profile with $R=1$ (equation \ref{eq:psi_eq0}).

The droplet spreads due to unbalanced Young stress and equilibrates when it approaches the static angle $\theta_s$ in the case of a clean droplet. According to \cite{cai2014phase}, from mass conservation, the final equilibrium profiles are given as follows:
\begin{equation}    \label{eq:spreadingL}
    L = 2R_0\sqrt\frac{\pi}{2(\theta_s-\sin\theta_s\cos\theta_s)}\sin\theta_s,\;\;\;\;\;\;  H= R_0\sqrt\frac{\pi}{2(\theta_s-\sin\theta_s\cos\theta_s)}(1-\cos\theta_s) \;
\end{equation}

We show the analytical and simulation values of $L$ and $H$ in figure \ref{fig:sl} for different wettablilty ranging from $45^\circ$ to $135^\circ$. The results show good agreement with the analytical solution. Having confirmed that the model reproduces clean droplet spreading, we now turn to the effect of soluble surfactants.
\begin{figure}
  \centerline{\includegraphics[width=0.6\textwidth]{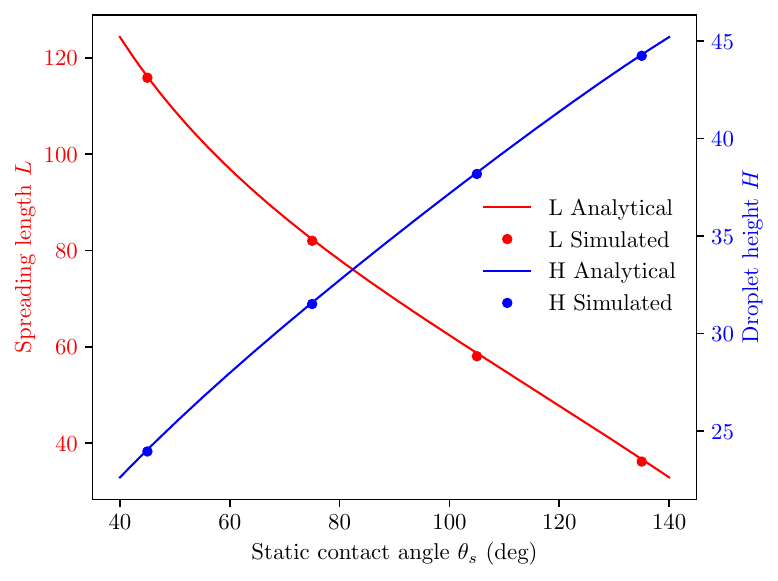}}
  \caption{Comparison of analytical and numerical values of spreading length $L$ and droplet height $H$ at different static angles $\theta_s$}
\label{fig:sl}
\end{figure}

\begin{figure}
  \centerline{\includegraphics[width=1\textwidth]{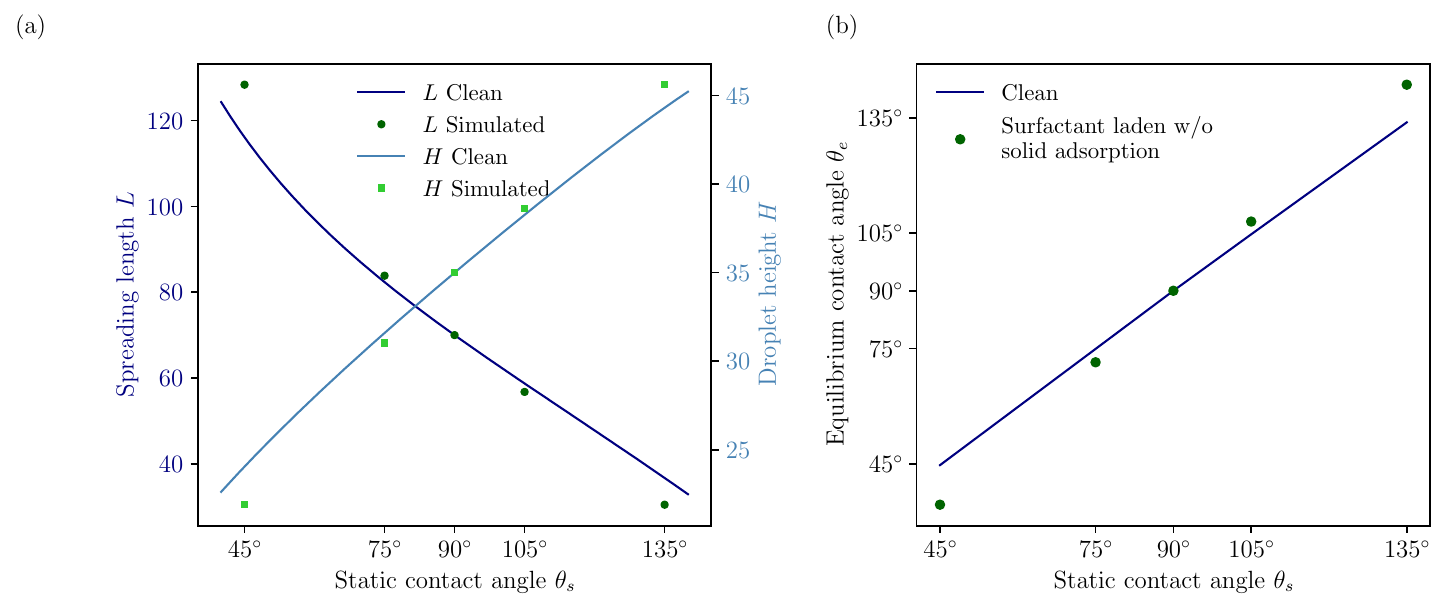}}
  \caption{(a) Comparison of clean analytical and surfactant laden (without solid adsorption) numerical values of spreading length $L$ and droplet height $H$ at different static angles $\theta_s$ (b) Comparison of clean analytical and surfactant laden (without solid adsorption) numerical values of different equilibrium angles $\theta_e$ at different static angles $\theta_s$ }
\label{fig:sl-surf}
\end{figure}


\subsection{Soluble surfactants without wall adsorption}
\label{sec:surfactant_no_wall}

We now study the spreading droplet case with soluble surfactants, but without modelling solid surface adsorption. This case reproduces the work of \cite{zhu2019surfmain}. We compare the equilibrium state of a clean droplet to a surfactant-laden droplet. The Langmuir adsorption constant $\psi_c$ and the parameter Ex are 0.017 and 1 respectively. The value of Pi is calculated from equation \ref{eqbm-zhu}. Our model differs slightly from \cite{zhu2019surfmain} in the bulk free energy due to the addition of the surface smoothing term $|\nabla\psi|^2$.

Figure \ref{fig:sl-surf}b shows the effect of surfactants on equilibrium contact angles: hydrophilic substrates ($< 90^\circ$) become more hydrophilic, while hydrophobic substrates ($> 90^\circ$) become more hydrophobic when surfactants are added. Panel (a) demonstrates that the spreading length $L$ and droplet height $H$ can be calculated using the analytical spherical cap solution. Without solid adsorption, the droplet maintains a perfect spherical cap geometry in all cases.

This behavior arises from Young's equation with only fluid-fluid surface tension reduction:
\begin{equation}    
\sigma \cos\theta_e = \sigma_0\cos\theta_0
\label{eq:youngs_no_wall}
\end{equation}
where $\sigma_0$ and $\theta_0$ are the clean fluid-fluid surface tension and equilibrium angle, while $\sigma$ and $\theta_e$ are the corresponding surfactant-laden values. The reduced interfacial tension $\sigma < \sigma_0$ rescales Young's equation, amplifying the existing wetting character: hydrophilic substrates amplify their hydrophilicity; hydrophobic substrates amplify their hydrophobicity. This rescaling behavior has been widely accepted in the numerical literature \citep{lai2010numerical, xu2014level, zhang2014derivation}. \cite{bera2018counteracting} show that there are a number surfactants that do not change the wettability of the solid: they give the same contact angle as a simple liquid with the same liquid–vapor surface tension. This model is also valid if the solid-liquid and solid-vapour surface tensions change by the same amount.

However, this prediction contradicts most experimental observations. Across diverse surfactant-substrate systems, experiments consistently show that surfactants reduce contact angles systematically-both hydrophilic and hydrophobic substrates shift toward increased hydrophilicity, not amplified wetting character. This discrepancy reveals a critical limitation: models that neglect solid surface adsorption capture only the fluid-fluid interfacial effect and cannot explain the experimentally observed universal trend. To resolve this, we now incorporate solid surface adsorption.

\subsection{Soluble surfactants with wall adsorption $(\beta_{sl} > \beta_{sg})$}
\label{sec:modelsolads}
We now finally incorporate solid adsorption of surfactants into the model and study the spreading length, droplet height and equilibrium angles. The coefficient of the square gradient term, $\lambda_s$ is taken to be $\lambda/2$. The strength of surfactant $\beta_{sl}$ on the solid-droplet interface is taken to be 0.9 and the strength of surfactant $\beta_{sg}$ on the solid-ambient interface is taken to be 0.01. Also the solid-droplet surface tension $\sigma_{sl}$, and the solid-ambient surface tension $\sigma_{sg}$ are taken such that the equilibirum angle for the clean droplet is corresponding to the case we want to study, in accordance to Young's equation. For example, we take $\sigma_{sg}=0.0107/1.5$ N/m, $\sigma_{sl}=0.0107/2.5$ N/m and $\sigma_{sg}=0.0107$ N/m for simulating a spreading of a surfactant laden droplet with a clean case equilibrium angle of $75^\circ$. 

\begin{figure}
  \centerline{\includegraphics[width=0.7\textwidth]{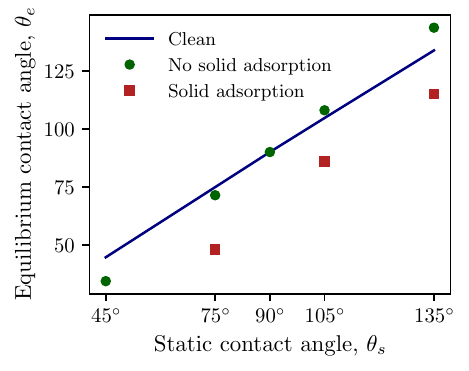}}
  \caption{Comparison of clean and surfactant laden droplet with and without solid adsorption values of equilibrium angle $\theta_e$ at different static angles $\theta_s$}
\label{fig:theta-all}
\end{figure}
\begin{figure}
  \centerline{\includegraphics[width=\textwidth]{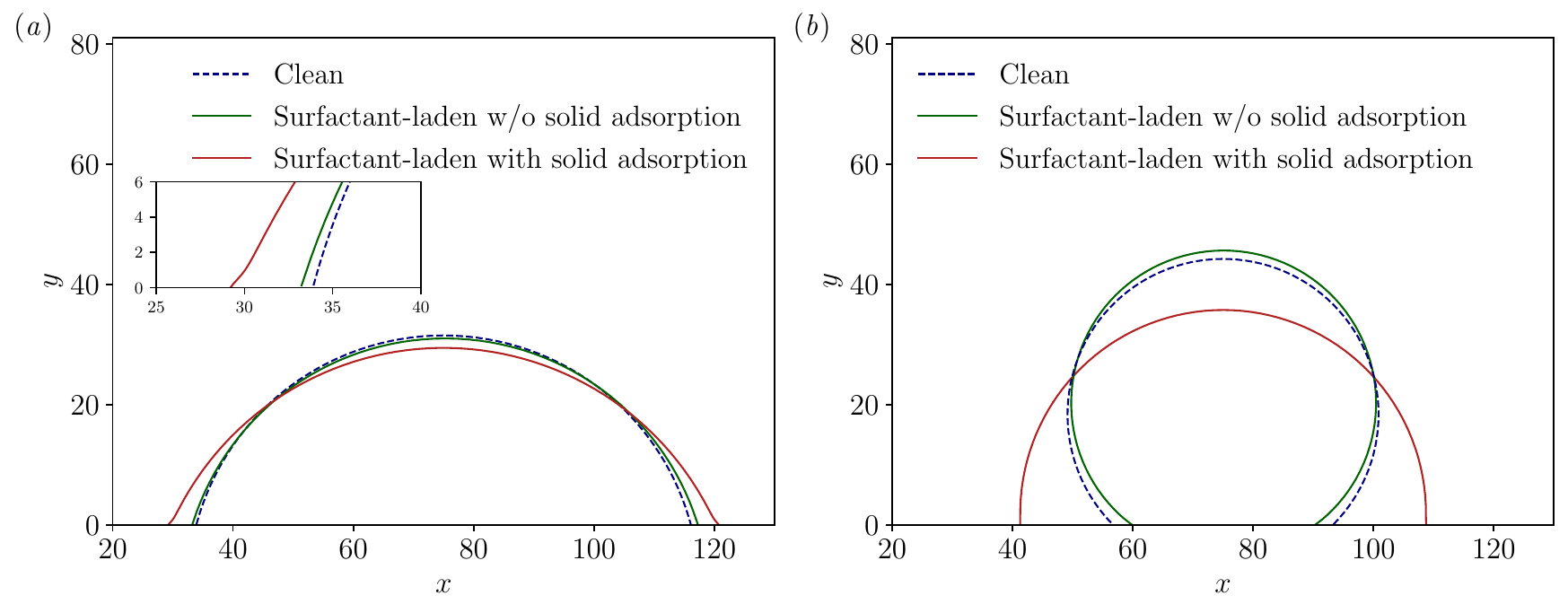}}
  \caption{Comparison of clean and surfactant laden droplet with and without solid adsorption steady states on a a) hydrophilic substrate (with $\theta_e$ of clean droplet being $75^\circ$) b) hydrophobic substrate (with $\theta_e$ of clean droplet being $135^\circ$). The inset in the left figure shows the bending of interface near the contact line when solid adsorption is added. }
\label{fig:contour-compare}
\end{figure}
\begin{figure}
  \centerline{\includegraphics[width=1\textwidth]{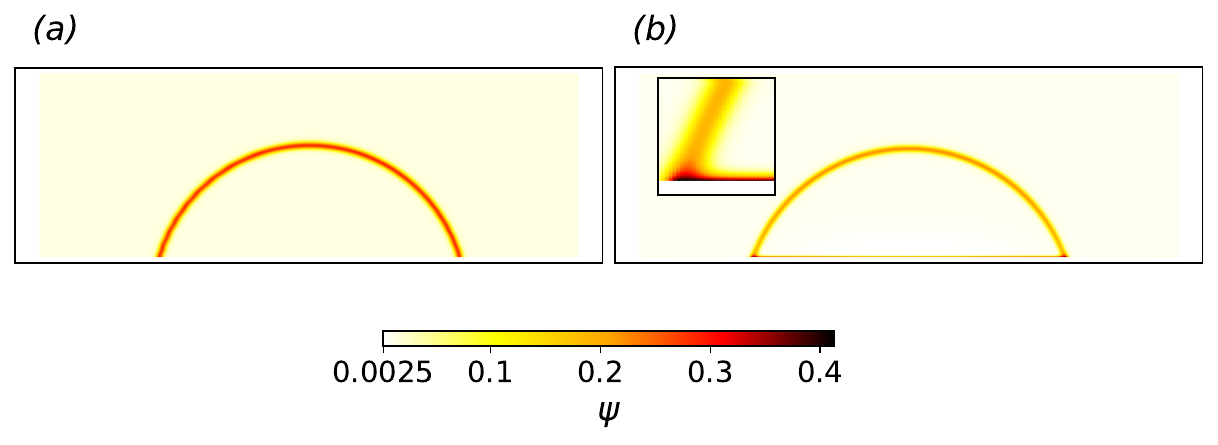}}
  \caption{ Surfactant concentration of the case set up shown in figure \ref{fig:contour-compare} with $\theta_e$ of clean droplet being $75^\circ$ (a) at steady state with only interfacial adsorption b) at steady state with interfacial and solid adsorption. The inset shows the surfactant concentration near the contact line with wall adsorption.  }
\label{fig:surf-concentration}
\end{figure}

We observe from Figure \ref{fig:theta-all} that regardless of the wettability of the substrate, the apparent equilibrium contact angle always is smaller than the clean case apparent equilibrium angle when surfactant adsorption on solid is considered i.e. the substrate becomes more hydrophilic. This finding is also in line with experiments in general \citep{kwok1999, johnson1986effects} and as shown in figure \ref{fig:exp-st} where a reduction in contact angle is observed as PEO and PEG are added to water. This emphasizes the importance of including the effect of the solid-fluid surface energy reduction and the effect it has on a molecular level in modelling surfactant laden moving contact lines problems. We rewrite the equality analogous to equation \ref{eq:youngs_no_wall} here for completeness:
\begin{equation}    \label{eq:modyoungseq2}
\begin{aligned}
\sigma \cos\theta_{e} &=  \sigma_0 \cos\theta_{0} +  \sigma_{w10}\beta_{s1}\log(1-\psi) -  \sigma_{w20}\beta_{s2}\log(1-\psi) \;
\end{aligned}
\end{equation}
where $\sigma_0$ and $\theta_0$ are the surface tension between the fluids and equilibrium angle of the clean case, and $\sigma$ and $\theta_e$ are the surface tension between the fluids and equilibrium angle of the surfactant laden case. Here the parameters are chosen such that the surface tension between the droplet and solid decreases a lot more than the surface tension between the ambient fluid and solid.

The steady state droplet contours for the hydrophilic and hydrophobic surfaces are presented in Figure \ref{fig:contour-compare}. The inset in panel \ref{fig:contour-compare} (a) reveals a subtle but important feature: the interface near the contact line exhibits a slight bending when solid adsorption is included (red curve) compared to the cases without solid adsorption. This bending arises from the competition between Marangoni stresses and Laplace pressure near the contact line. 

The physical mechanism is as follows: solid adsorption creates a surfactant concentration gradient near the wall (visible in Figure~\ref{fig:surf-concentration} (b)), which generates a surface tension gradient $\nabla\sigma$ along the fluid--fluid interface directed away from the contact line. This gradient imposes a tangential Marangoni stress at the interface, driving flow in the adjacent bulk fluid. Near the contact line, the fluid--fluid interface meets the solid wall at a finite angle, 
geometrically confining the Marangoni-driven flow. By the incompressibility constraint $\nabla \cdot \mathbf{u} = 0$, this geometric confinement generates a local pressure field in the bulk fluid adjacent to the contact line. It is this pressure field that 
modifies the normal stress jump across the interface, and the interface responds by adjusting its local curvature $\kappa$ until the normal stress balance $\Delta P = \sigma\kappa$ is satisfied. The interface bending visible in the inset of figure~\ref{fig:contour-compare}(a) is the geometric signature of this curvature adjustment. This sequential coupling --- Marangoni stress, bulk flow, pressure generation via incompressibility, and curvature adjustment --- has been described by \citet{chan2005surfactant} and observed experimentally by \citet{gokhale2005spreading} and \citet{leite2018enhanced}. This coupling between surfactant distribution, Marangoni stress, and interface curvature represents a key feature of surfactant-laden contact line dynamics that models neglecting solid adsorption cannot capture, as shown in figure \ref{fig:surf-concentration} a).

In short, for hydrophilic surfaces, surfactant-induced spreading is primarily driven by fluid-fluid interface adsorption, as the reduced interfacial tension directly lowers the equilibrium contact angle. Conversely, on hydrophobic surfaces, solid-droplet interface adsorption becomes critical: merely reducing fluid-fluid tension would increase the angle, so decreasing the solid-droplet surface tension is necessary to achieve spreading. This mechanism has been confirmed through MD simulations by \cite{staniscia2022tuning}.



\subsection{Autophobing: antisurfactant effect through reversed asymmetric surface adsorption}

Autophobing is a phenomenon wherein droplets exhibit increased contact angles upon surfactant addition, contrary to typical expectations that surfactants promote wetting. In its most striking form, droplets initially spread upon deposition but then spontaneously retract to a larger equilibrium contact angle-a dynamic phenomenon reported by \citet{bera2016surfactant} and \citet{tadmor2019drops}. Experimentally, autophobing occurs when surfactants preferentially adsorb on the solid-ambient interface via a precursor film, reducing that surface tension more than the solid-droplet interface, thereby destabilizing the initial wetting state \citep{bera2016surfactant, bera2021antisurfactant}. In contrast to \S\ref{sec:modelsolads} , which considered strong adsorption on the liquid side ($\beta_{sl} \gg \beta_{sg}$), we now examine the reversed asymmetry ($\beta_{sg} \gg \beta_{sl}$). This demonstrates the versatility of our wall adsorption mechanism and its capacity to capture autophobing-a phenomenon that prior models neglecting solid surface adsorption could not predict.

The autophobing effect emerges through the modified Young's equation:
\begin{equation}
\sigma \cos \theta_e = \sigma_0 \cos \theta_0 + \sigma_{sg,0}\beta_{sg} \log(1 - \psi) - \sigma_{sl,0}\beta_{sl} \log(1 - \psi)
\label{eq:autophobing_young}
\end{equation}
When $\beta_{sg} \gg \beta_{sl}$, surfactants preferentially accumulate at the solid-ambient interface, producing substantial surface tension reduction:
\begin{equation}
\Delta\sigma_{sg} = -\sigma_{sg,0}\beta_{sg} \log(1 - \psi) \gg \Delta\sigma_{sl} = -\sigma_{sl,0}\beta_{sl} \log(1 - \psi).
\label{eq:asymmetric_reduction}
\end{equation}
The asymmetric reduction-with the ambient side experiencing dramatically greater surface tension suppression-dominates Young's equation, shifting the equilibrium contact angle toward higher values regardless of initial wettability. This is the inverse of the typical spreading case in \S 4.3, where $\beta_{sl} > \beta_{sg}$ produces hydrophilicity.

To investigate autophobing, we configure the model with reversed adsorption asymmetry: $\beta_{sg} = 0.9$ (strong adsorption on solid-ambient interface) and $\beta_{sl} = 0.01$ (weak adsorption on solid-droplet interface). This represents a scenario where surfactant molecules preferentially bind to the high-energy solid-ambient interface. We test two initial wettability states to demonstrate the universality of wetting reversal: a hydrophilic substrate with $\theta_0 = 75^\circ$ and a hydrophobic substrate with $\theta_0 = 105^\circ$. All other parameters remain consistent with \S\ref{sec:modelsolads}. Critically, bulk surfactant solubility remains symmetric across phases ($\psi_b$ identical), confirming that autophobing emerges purely from wall adsorption asymmetry. 

\begin{figure}
  \centerline{\includegraphics[width=0.7\textwidth]{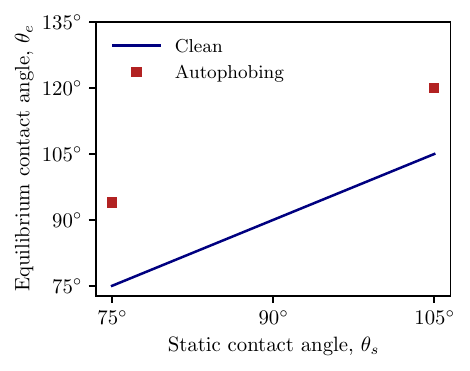}}
  \caption{Comparison of clean and surfactant laden droplet values of equlibrium angle $\theta_e$ at different static angles $\theta_s$ in the process of autophobing}
\label{fig:auto}
\end{figure}
As shown in Figure \ref{fig:auto}, simulations reveal pronounced wetting reversal: the hydrophilic case shifts from $75^\circ$ to $94^\circ$ (a change of $\Delta\theta = +19^\circ$), while the hydrophobic case shifts from $105^\circ$ to $120^\circ$ ($\Delta\theta = +15^\circ$) as shown in figure~\ref{fig:auto}. These approximately $15^\circ$ contact angle increases match the experimental range reported by \citet{bera2016surfactant}, who observed autophobing-induced angle changes of 10--20\% for comparable surfactant systems. Critically, both substrates shift in the hydrophobic direction. This is qualitatively distinct from the standard rescaling of Young's law without wall adsorption, and from the typical spreading case in \S\ref{sec:modelsolads}, where it becomes more wetting. The reversed adsorption asymmetry reverses the wetting trend entirely. 
\begin{figure}
  \centerline{\includegraphics[width=0.8\textwidth]{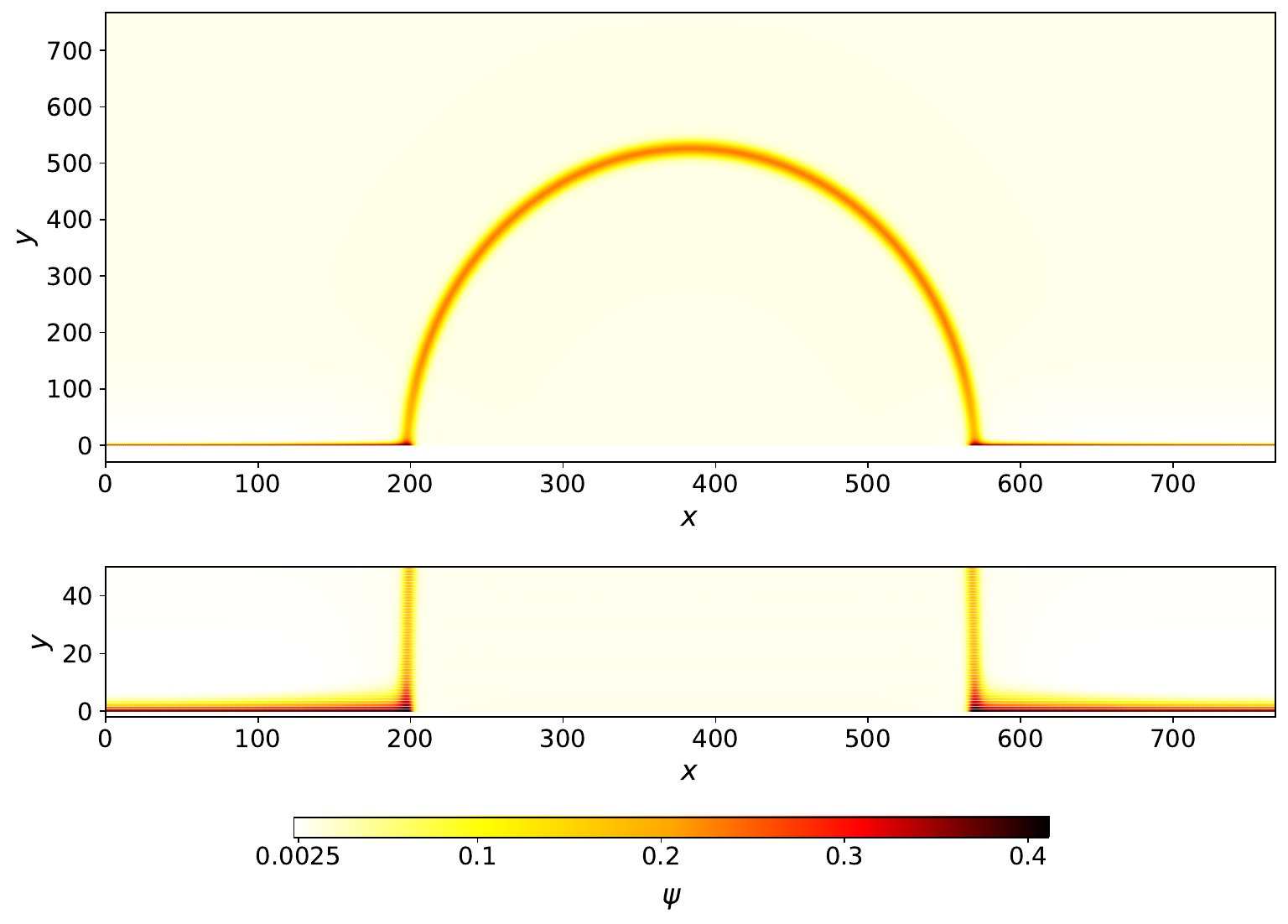}}
  \caption{Steady state of a surfactant laden droplet (clean case equilibrium angle being $75^\circ$) which has undergone autophobing equilibrated at $94^\circ$. The surfactant accumulation at the wall outside the droplet is observed. The region near the contact line has been zoomed in the bottom figure.}
\label{fig:autophobing_conc}
\end{figure}
Figure \ref{fig:autophobing_conc} visualizes the surfactant concentration field at equilibrium and reveals the physical basis for autophobing. The image shows a striking red adsorption band localized at the solid-ambient fluid interface (outer wall, corresponding to the 'dry' substrate ahead of the contact line), in sharp contrast to negligible accumulation at the solid-droplet interface (inner wall, corresponding to the 'wet' substrate under the droplet). The wall surfactant concentration reaches $\psi_{\text{wall}} \approx 0.41$ on the ambient fluid side, compared to approximately $0.011$ on the droplet side. The adsorption layer thickness on the ambient fluid side is $\delta_{sg} \approx 1.4\epsilon$, whereas the droplet side shows minimal accumulation with $\delta_{sl} \approx 0.5\epsilon$.

This pronounced asymmetry in surface adsorption directly translates to differential surface tension modification via the Langmuir--Szyszkowski equation. The ambient fluid side experiences a surface tension reduction of approximately 47.5\%, while the droplet side sees only 0.01\% reduction. To confirm the mechanism, we employ these measured wall concentrations in Young's equation \eqref{eq:autophobing_young}. We take the effective surface tension in the fluid--fluid interface to be $0.9$ times the clean interfacial tension according to the figure \ref{fig:isotherm_validation}  in Appendix \ref{ap:flfl}. Computing the predicted contact angles yields $\theta_{\text{calc}} = 92.4^\circ$ for the hydrophilic case (compared to simulated $94^\circ$) and $\theta_{\text{calc}} = [120.1]^\circ$ for the hydrophobic case (compared to simulated $120^\circ$). The close agreement validates that the model correctly captures the autophobing mechanism: asymmetric wall adsorption produces differential surface tension modification, which determines the equilibrium wetting state via Young's equation. The mechanism is mathematically transparent-Young's equation itself reveals why reversed asymmetry must produce reversed wetting behavior-and robust to parameter variation.

 While our model captures autophobing in sessile droplet geometries, the phenomenon has also been extensively studied in thin film configurations where Marangoni flows and film stability play additional roles \citep{craster2007autophobing}. 

\begin{figure}
  \centering
  \includegraphics[width=0.5\textwidth]{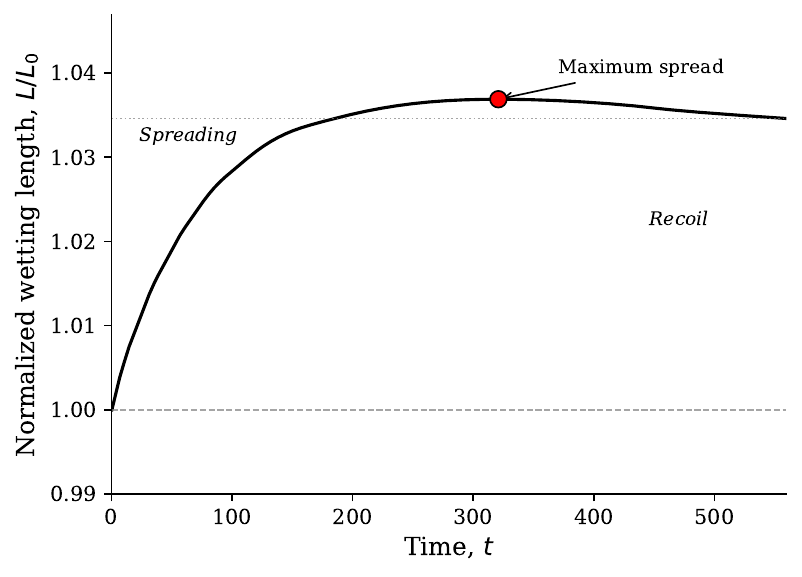}
  \caption{Temporal evolution of the normalized wetting length $L/L_0$ for the autophobing case ($\beta_{sg} = 0.9$, $\beta_{sl} = 0.01$) with clean equilibrium angle $\theta_0 = 75^\circ$. The droplet initially spreads due to unbalanced Young stress, reaching maximum extent at $t \approx 321$ (red marker). As surfactant accumulates at the solid-ambient fluid interface ahead of the contact line, the equilibrium shifts toward higher contact angles and the droplet recoils. The dashed line indicates the initial wetting length $L_0$, while the dotted line marks the final equilibrium value. This transient behaviour-initial spreading followed by spontaneous retraction-is the hallmark of dynamic autophobing observed experimentally \citep{bera2021antisurfactant}}
  \label{fig:autophobing_dynamics}
\end{figure}

\subsubsection{Connection to transient autophobing dynamics}
The kinetic boundary condition (equation \ref{eq:ads-bc}) with parameter $\Gamma_s$ determines how rapidly surfactants adsorb on the wall, thereby controlling the temporal evolution of the contact line during spreading. We have chosen $\Gamma_s=1000$, which gives us the evolution for the spreading radius as shown in Figure \ref{fig:autophobing_dynamics}. Our model can capture the transient autophobing dynamics observed experimentally: the droplet initially spreads because surfactant has not yet accumulated outside the contact line, but as surfactant progressively adsorbs on the solid-ambient interface (the ``dry'' wall ahead of the contact line), the contact angle increases and the droplet retracts \citep{lee2008kinetics}. This initial spreading followed by spontaneous retraction is the hallmark of dynamic autophobing reported in experiments \citep{bera2016surfactant, tadmor2019drops}.  We mainly focus on the equilibrium state in this paper, while transient autophobing trajectories remains for future work with time-resolved experimental data.

Additionally, we assume identical bulk surfactant solubility $\psi_b$ in both phases, whereas experimentally surfactants may preferentially dissolve in one phase. In the dilute regime considered here ($\psi_B<<1$), differential solubility between phases manifests primarily as a rescaling of the effective surfactant concentration in one phase, and does not introduce fundamentally new coupling physics between bulk transport and wall adsorption. The equilibrium wetting behaviour is therefore qualitatively preserved under this assumption. If greater fidelity is required, differential bulk solubility can in principle be incorporated by modifying the bulk free energy through the parameter $F_\text{Ex}$, and this extension is left for future work.

The robustness of autophobing is evident from equation \eqref{eq:autophobing_young} itself. The contact angle depends explicitly on the difference in adsorption-weighted surface tension changes. Thus autophobing is robust: it is a direct mathematical consequence of reversed asymmetry in elasticity parameters, not a sensitive tuning of the model.


\subsection{Verification of solid adsorption scaling laws}
\label{sec:scaling_laws}

\begin{figure}
  \centerline{\includegraphics[width=0.5\textwidth]{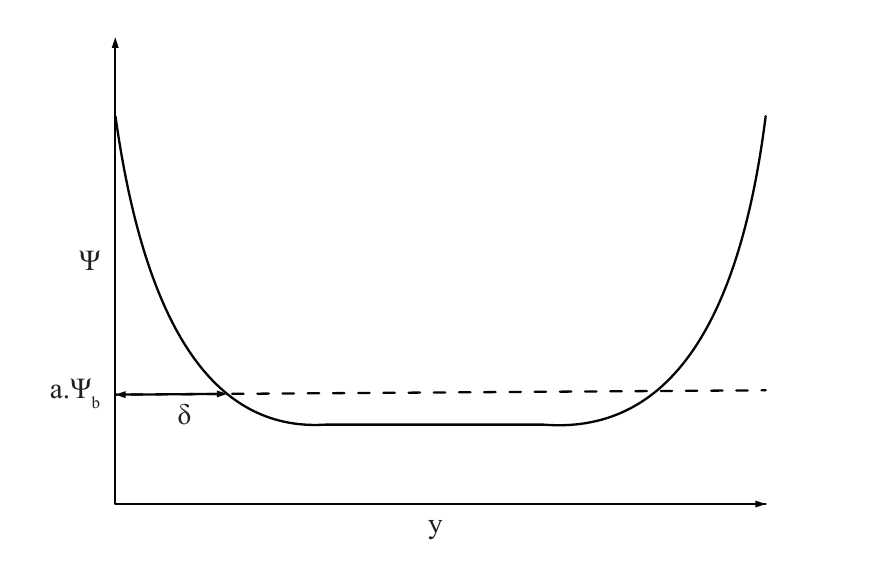}}
  \caption{Illustration of a surfactant-laden single-phase system with walls on both ends at steady state}
\label{fig:setadysinglephase}
\end{figure}

\begin{figure}
  \centerline{\includegraphics[width=1.0\textwidth]{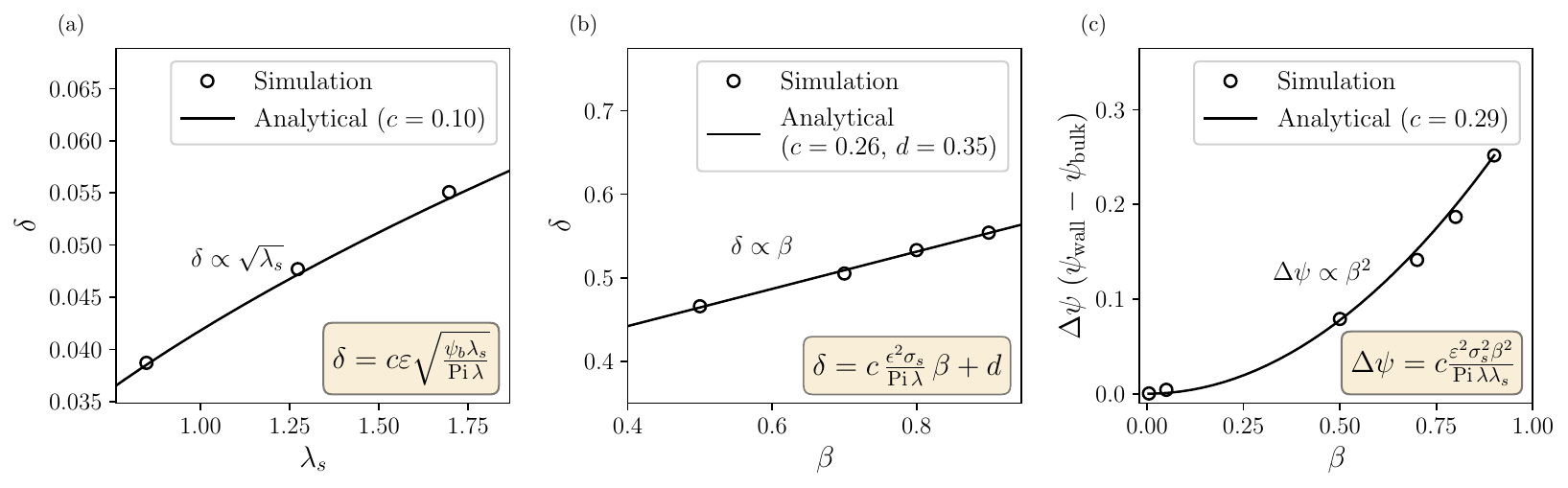}}
  \caption{a) Scaling of $\delta$ wrt $\lambda_s$ in Regime 1 b) linear scaling of $\delta$ wrt $\beta$ in Regime2 c) quadratic scaling of $\Delta \psi$ with respect to $\beta$ across the entire range of $\beta$ irrespective of the regime. The value of coefficient $a$ for Regime 1 and Regime 2 are 0.9 and 1.0 respectively.  }
\label{fig:delta-bs}
\end{figure}

To validate the scaling predictions derived in \S\ref{surf-bl}, we perform a systematic numerical study using a 1D single-phase system with walls on the top and bottom and a length of $L=100$.  For Regime 1, we vary the diffusivity parameter $\lambda_s$ from $20$ to $40$ times $\lambda$, while fixing $\beta = 0.03$, $\text{Pi} = 0.1841$, and $\text{Cn} = 0.01$.  For Regime 2, we vary the adsorption strength parameter $\beta$ from 0.5 to 0.9, while fixing $\lambda_s = \lambda/2$, and keeping Pi, and Cn unchanged. We define the adsorption layer thickness $\delta$ as the distance from the wall at which $\psi = a \cdot \psi_b$ (illustrated in Figure \ref{fig:setadysinglephase}), and the concentration excess $\Delta\psi$ as the difference between wall and bulk surfactant concentrations.

The simulations reveal two distinct scaling regimes for the adsorption layer. In Regime 1 ($\Delta\psi \ll \psi_b$), where wall affinity is weak, the adsorption layer thickness scales with the square root of the surfactant diffusivity: $\delta \propto \sqrt{\lambda_s}$ as shown in Figure \ref{fig:delta-bs} (a), where $\delta$ is defined as the distance from the wall at which $\psi = a\cdot\psi_b$ with $a = 0.9$ (Figure \ref{fig:setadysinglephase}). This square-root dependence indicates that increased surfactant diffusivity allows surfactant molecules to penetrate further into the bulk in the direction normal to the wall, resulting in a thicker adsorption layer.

In Regime 2 ($\Delta\psi \gg \psi_b$), where wall affinity is strong, the layer thickness scales linearly with surfactant strength: $\delta \propto \beta$ as shown in figure \ref{fig:delta-bs} (b), where $\delta$ is defined as the distance from the wall at which $\psi = a\cdot\psi_b$ with $a = 1.0$ (Figure \ref{fig:setadysinglephase} (a)). This linear relationship reflects the stronger driving force for accumulation when wall affinity is high: surfactant adsorbs more readily and the layer extends further into the bulk with a steeper concentration gradient than in Regime 1.

Additionally, the concentration excess $\Delta\psi = \psi_{\text{wall}} - \psi_{\text{bulk}}$ follows a quadratic dependence on $\beta$ across all regimes: $\Delta\psi \propto \beta^2$ (Figure \ref{fig:delta-bs}c). These scaling behaviors are in excellent agreement with the theoretical predictions from \S\ref{surf-bl} (Equations \ref{eq:scaling3}, \ref{eq:scaling4} and \ref{eq:scaling2.5}) with the constant of proportionality $c$ being within an order of magnitude, validating that the model correctly captures the equilibrium structure of adsorbed surfactant layers at solid--fluid interfaces.

 \subsection{Motion of surfactant laden droplet enclosed in a Couette flow}
\label{sec:pl-pois}

In this subsection we introduce a background shear flow to the problem and simulate a surfactant laden droplet in a Couette flow on a solid substrate. With this, we intend to get a detailed picture of the molecular effect i.e. change in equilibrium angle by adding solid adsorption and hydrodynamic effects such as Marangoni stresses that result near the contact line and along the fluid-fluid interface. 

  \begin{figure}
  \centerline{\includegraphics[width=0.8\textwidth]{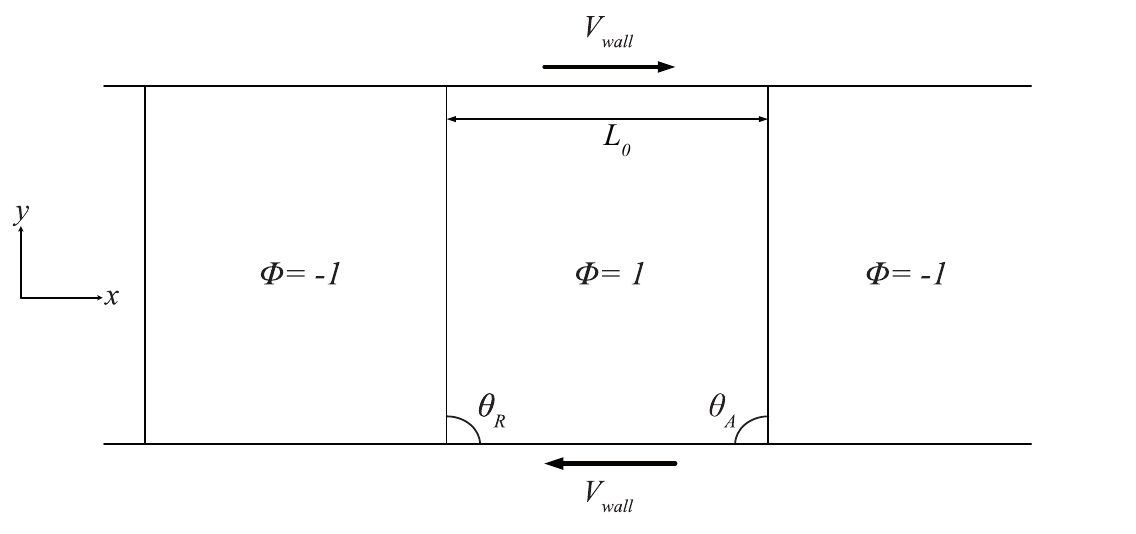}}
  \caption{Schematic of a two-phase couette flow with walls moving in opposite directions. The advancing and receding angles are labelled. }
\label{fig:pla-poise}
\end{figure}

\subsubsection{Setup and Parameters}

To investigate droplet stability under shear flow, we simulate a surfactant-laden droplet confined between two parallel walls in relative motion. The configuration consists of a droplet of thickness $0.4H$ confined between walls separated by distance $H$, where the top wall moves with velocity $V_{wall>0}$ (rightward) and the bottom wall moves with velocity $V_{wall}<0$ (leftward), creating a planar Couette flow as shown in figure \ref{fig:pla-poise}. The computational domain has length $4H$ in the streamwise direction with periodic boundary conditions. The two fluids are density and viscosity matched.

The simulations are performed at Reynolds number $\text{Re} = 0.1$ and Cahn number $\text{Cn} = 0.05$. Following \citet{Yue2020cah}, we define $\Pi=0.5$, $S_\phi=0.01$ which contains $M_\phi$, and $S_\psi=0.01\sqrt{10}$ which contains $m_\psi$. The clean droplet equilibrium contact angle is $\theta_{0} = 80°$, which is set by choosing $\sigma_{lg} = 0.0107$ N/m, $\sigma_{sl} = 0.0107/3$ N/m, and $\sigma_{sg} = 0.0107/2$ N/m.

 \begin{figure}
  \centerline{\includegraphics[width=1\textwidth]{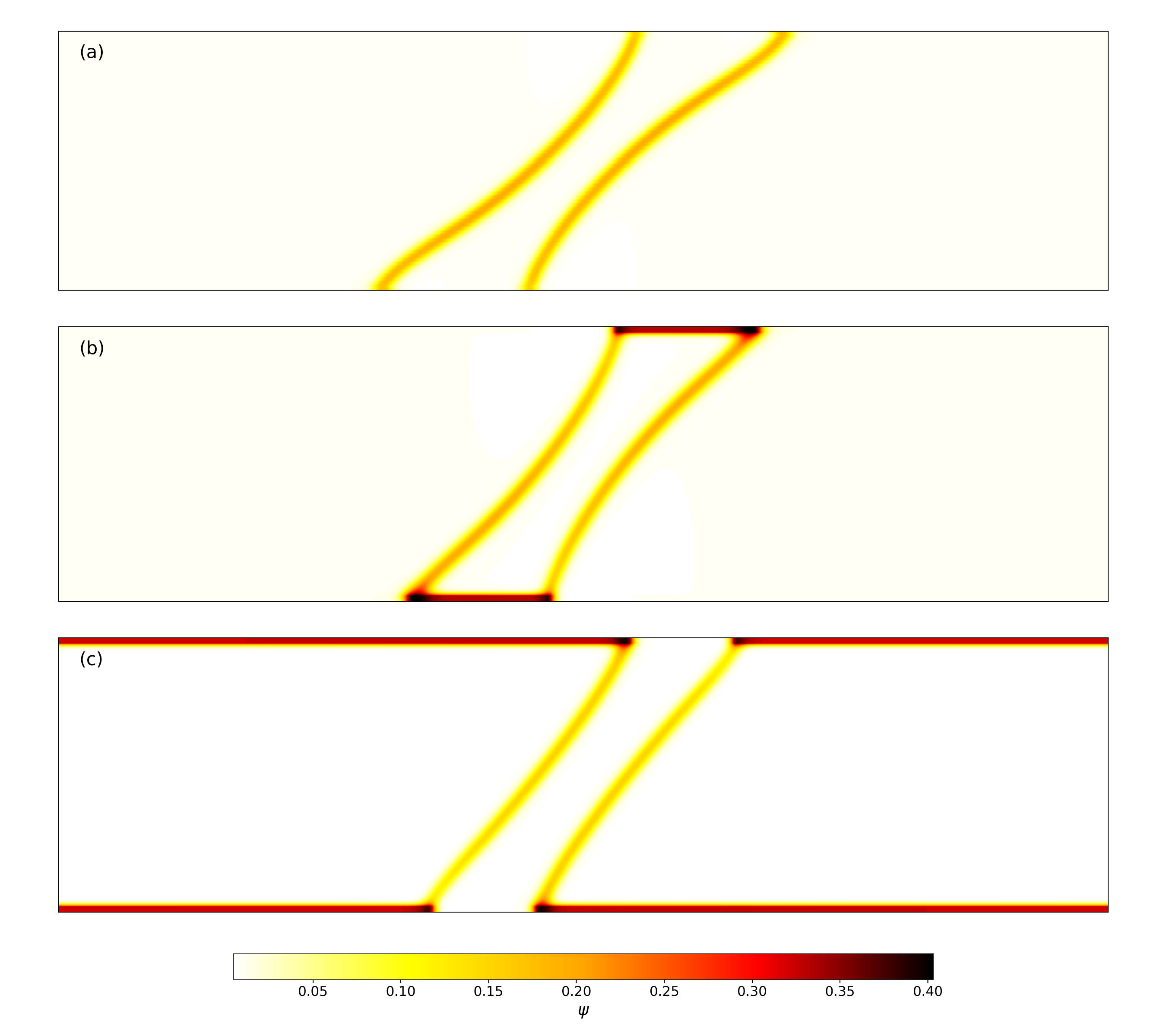}}
  \caption{Surfactant concentration at steady state for (a) no solid adsorption (b) enhanced wetting ($B_{sl}=0.9$) (c) autophobing ($B_{sg}=0.9$) cases. This is for $\text{Ca}=0.02$. }

\label{fig:couettestd}
\end{figure}
\begin{table}
  \centering
  \caption{Summary of surfactant configurations for the Couette flow simulations. The approximate equilibrium contact angle $\theta_{eq}$ for surfactant laden cases are measured using the modified Young equation (eq \ref{eq:modyoungseq2}). The adhesion force $(\sigma_{sg} - \sigma_{sl})$ quantifies the mechanical resistance to contact line motion. The critical capillary number $Ca_{crit}$ denotes the threshold above which droplet breakup occurs.}
  \label{tab:couette_summary}
  \begin{tabular}{lccccc}
    \hline
    Configuration & $\beta_{sl}$ & $\beta_{sg}$ & $\theta_{eq}$ ($^\circ$) & $\sigma_{sg} - \sigma_{sl}$ & $Ca_{crit}$ \\
    \hline
    Clean & --- & --- & 80 & 0.00178 & 0.02--0.03 \\
    No solid adsorption & 0 & 0 & $\approx$78 & 0.00178 & 0.02--0.03 \\
    Enhanced wetting & 0.9 & 0.01 & $\approx$55 & 0.00293 & 0.03--0.04 \\
    Autophobing & 0.01 & 0.9 & $\approx$95 & $-$0.00071 & 0.02--0.03 \\
    \hline
  \end{tabular}
\end{table}
The steady state of the surfactant laden cases are given in figure \ref{fig:couettestd}. Before examining the detailed dynamics, we summarize the four configurations studied in table~\ref{tab:couette_summary}. Each configuration represents a distinct regime of surfactant--wall interaction: (1) the clean case serves as the baseline with no surfactant present; (2) the no solid adsorption case ($\beta_{sl} = \beta_{sg} = 0$) isolates the effect of interfacial tension reduction without wall chemistry modification; (3) the enhanced wetting case ($\beta_{sl} = 0.9$, $\beta_{sg} = 0.01$) represents strong preferential adsorption at the solid-droplet interface; and (4) the autophobing case ($\beta_{sl} = 0.01$, $\beta_{sg} = 0.9$) represents the reversed asymmetry with strong adsorption at the solid-ambient interface. The key distinction between these configurations lies not only in the equilibrium contact angle, but also in the adhesion force $(\sigma_{sg} - \sigma_{sl})$. As we show below, the adhesion force-rather than the contact angle magnitude-controls the critical capillary number for droplet breakup. Here, the subscripts ``sl'' and ``sg'' denote the solid interface with the droplet phase (inside the droplet) and the ambient phase (outside the droplet), respectively. The capillary number $\text{Ca} = \mu V_{wall}/\sigma_0$ is systematically varied to identify the critical value $\text{Ca}_\text{crit}$ at which droplet breakup occurs.

\subsubsection{Steady-State Contact Angles}

 \begin{figure}
  \centerline{\includegraphics[width=1\textwidth]{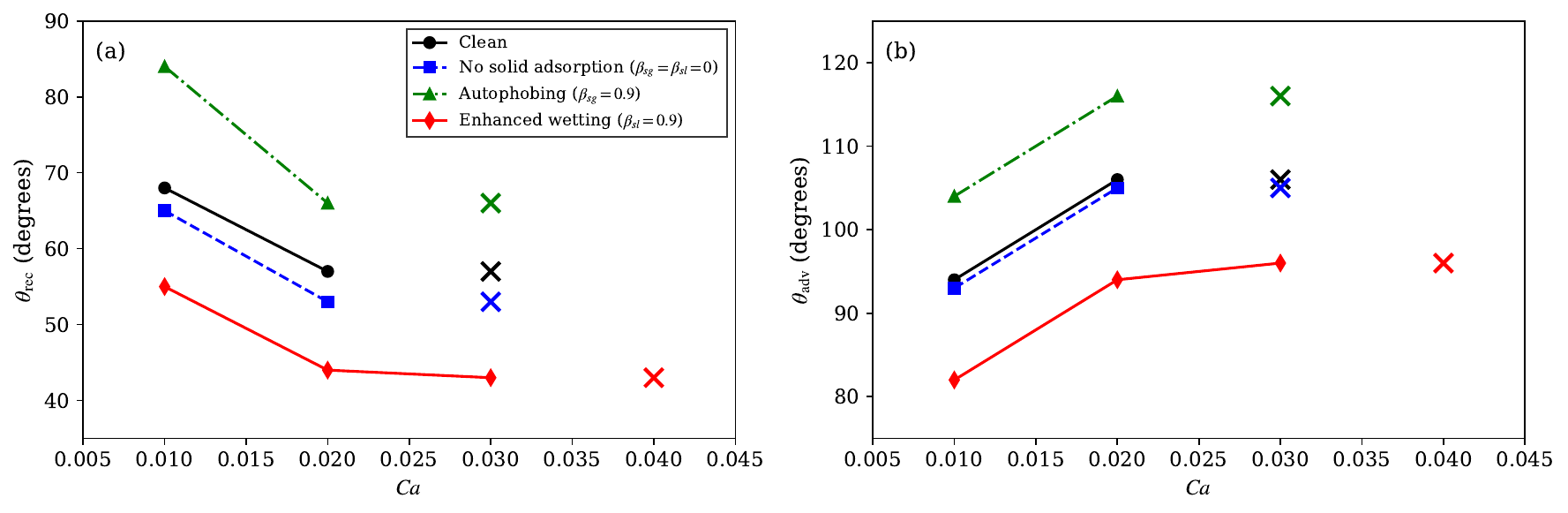}}
  \caption{Steady-state contact angles versus capillary number for (a) receding and (b) advancing contact lines. Enhanced wetting (red, $\beta_{sl}=0.9$) exhibits the lowest angles, followed by no solid adsorption (blue, $\beta_{sg}=\beta_{sl}=0$), clean (black), and autophobing (green, $\beta_{sg}=0.9$). At $\text{Ca} = 0.01$ and $0.02$, all configurations reach steady states. At $\text{Ca} = 0.03$, only enhanced wetting remains stable; clean, no solid adsorption, and autophobing undergo droplet breakup ($\times$ marks). All cases fail at $\text{Ca} = 0.04$. }
\label{fig:couette}
\end{figure}

Figure \ref{fig:couette} compares steady-state contact angles across all four configurations at $\text{Ca} = 0.01$ 
and $0.02$. The clean droplet has a hydrophilic equilibrium angle ($\theta_0 = 80^\circ$), and all observed 
trends are consistent with the equilibrium wetting behavior established in \S4.3. As capillary number increases, 
receding angles decrease while advancing angles increase, reflecting intensifying flow-induced distortion. The asymmetry between advancing and receding contact angles stems from non-equilibrium dynamics and should be distinguished from contact angle hysteresis-which describes this angular difference specifically at the onset of contact line motion.

The four cases maintain consistent ordering: enhanced wetting ($\beta_{sl} = 0.9$) exhibits the lowest angles, 
followed by no solid adsorption, clean, and autophobing ($\beta_{sg} = 0.9$), directly reflecting their 
equilibrium wetting tendencies established by surfactant adsorption.

The clean and no solid adsorption cases show remarkably similar behavior---differing by only 
$1^\circ$ on the advancing side and $3$--$4^\circ$ on the receding side. Without solid adsorption, surfactants 
reduce only $\sigma_{lg}$, leaving $\sigma_{sg}$ and $\sigma_{sl}$ unchanged. The modest interfacial tension reduction 
($\sigma_{lg} \approx 0.9\sigma_0$) produces small equilibrium angle shifts, with larger shifts at the receding 
contact line due to convective surfactant transport (Figure \ref{fig:couettestd} a). Critically, identical 
adhesion forces ($\sigma_{sg} - \sigma_{sl}$) yield nearly identical stability.

Enhanced wetting shows systematically lower contact angles due to strong adsorption at the 
solid-droplet interface ($\beta_{sl} = 0.9$), which reduces $\sigma_{sl}$ and increases the thermodynamic 
driving force for wetting. Both receding and advancing angles decrease symmetrically, 
indicating a fundamental equilibrium shift rather than flow-induced asymmetry.

Autophobing exhibits the opposite trend: strong adsorption at the solid-ambient interface 
($\beta_{sg} = 0.9$) lowers $\sigma_{sg}$, shifting equilibrium toward higher angles. At $\text{Ca} = 0.02$, the 
autophobing case maintains the highest receding ($66^\circ$) and advancing ($116^\circ$) angles. The 
receding angle drops $18^\circ$ from $\text{Ca} = 0.01$ to $0.02$-larger than other cases ($\sim11^\circ$)-while the 
advancing angle increases by $12^\circ$, similar to other configurations.

Viscous flow modulates these trends but preserves the fundamental ordering established by 
equilibrium wetting thermodynamics.

\subsubsection{Droplet Breakup at $\text{Ca} = 0.03$ and Unbalanced Young Stress}
\begin{figure}
  \centerline{\includegraphics[width=0.8\textwidth]{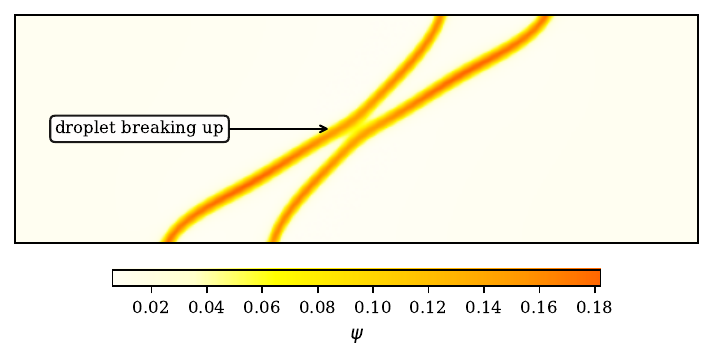}}
  \caption{Surfactant concentration of a no solid adsorption case at Ca=0.03 depicting the onset of droplet breakup after which the droplet disconnects into two.}

\label{fig:wet-fail}
\end{figure}
At $\text{Ca} = 0.03$, the distinction between configurations becomes decisive. Only the enhanced wetting case ($\beta_{sl} = 0.9$) maintains a stable steady state, achieving contact angles of $\theta_{rec} = 43°$ and $\theta_{adv} = 96°$. The clean, no solid adsorption, and autophobing configurations all undergo droplet breakup-the contact lines cannot maintain stable positions and the droplet breaks up. The onset of droplet breakup is shown in Figure \ref{fig:wet-fail}. This critical transition reveals that contact angle behavior alone does not determine stability under flow; rather, the resistance to contact line motion governed by the wall contribution to Young stress is the controlling factor.

To understand this transition, we examine the unbalanced Young stress at the contact line. At equilibrium, Young's equation balances the horizontal components of interfacial tensions:
\begin{equation}
\sigma_{lg} \cos\theta_{eq} = \sigma_{sg} - \sigma_{sl}
\end{equation}

Under flow, the contact line adopts a dynamic angle $\theta_{dyn}$ that deviates from equilibrium. This deviation changes $\cos\theta_{dyn}$, creating a non-zero unbalanced Young stress $\mathcal{L}$ \citep{qian2006variational}, which quantifies the net horizontal force per unit length acting at the contact line:
\begin{equation}
\mathcal{L} = \sigma_{lg} \cos\theta_{dyn} - (\sigma_{sg} - \sigma_{sl})
\end{equation}

 When $\mathcal{L} \neq 0$, there exists a net restoring force that drives the contact line toward its equilibrium position. The magnitude and direction of this restoring force depend on how much $\theta_{dyn}$ deviates from $\theta_{eq}$: if the contact line is dragged such that $\theta_{dyn}$ decreases, $\cos(\theta_{dyn})$ increases and $\mathcal{L}$ becomes positive (restoring force opposes the backward motion); conversely, if $\theta_{dyn}$ increases, $\cos\theta_{dyn}$ decreases and $\mathcal{L}$ becomes negative (restoring force opposes the forward motion). Viscous stresses from the flow work against this restoring force. At low capillary numbers, the restoring force is sufficient to stabilize the contact line at some dynamic angle. However, as Ca increases, viscous forcing intensifies and requires larger deviations in $\theta_{dyn}$ to generate sufficient restoring force. Droplet breakup occurs when the viscous drag exceeds the maximum restoring force that the unbalanced Young stress can provide.

Critically, the maximum restoring force is fundamentally limited by the magnitude of $\sigma_{sg} - \sigma_{sl}$---the difference in solid-fluid interfacial tensions. This quantity, which we term the adhesion force, represents the wall's contribution to the Young stress balance and determines how strongly the contact line resists motion along the solid surface. A larger adhesion force allows larger deviations in $\theta_{dyn}$ (and thus larger changes in $\cos\theta_{dyn}$) before the unbalanced Young stress is exhausted, enabling the contact line to withstand higher viscous stresses before failure.

For the clean and no solid adsorption cases, the adhesion force is identical: $\sigma_{sg} - \sigma_{sl} = 0.00178$ for both configurations. Although the no solid adsorption case has reduced interfacial tension ($\sigma_{lg} \approx 0.9 \sigma_0$), the solid-fluid interfacial tensions remain unchanged, yielding the same adhesion force. Consequently, both configurations fail at the same critical capillary number ($\text{Ca}_\text{crit} \approx 0.02$--$0.03$). The near-identical contact angles observed at $\text{Ca} = 0.01$ and $0.02$ reflect this shared adhesion force, and their simultaneous failure at $\text{Ca} = 0.03$ confirms that adhesion force, not interfacial tension alone, controls the stability limit.

For the enhanced wetting case, strong adsorption at the solid-liquid interface reduces $\sigma_{sl}$ substantially, yielding an adhesion force of $\sigma_{sg} - \sigma_{sl} \approx 0.00293$--approximately 65\% larger than the clean case. This enhanced adhesion allows the contact line to generate greater restoring forces through the unbalanced Young stress, resisting higher viscous stresses before failure. As a result, the enhanced wetting configuration extends the critical capillary number to $Ca_{crit} \approx 0.03$--$0.04$, representing a 20--30\% increase in the stable operating range compared to clean and no solid adsorption cases.

For the autophobing case, strong adsorption at the solid-ambient interface reduces $\sigma_{sg}$, resulting in a negative adhesion force: $\sigma_{sg} - \sigma_{sl} \approx -0.00071$. From the unbalanced Young stress perspective, the sign of $(\sigma_{sg} - \sigma_{sl})$ determines whether displacement generates a restoring force (positive adhesion) or fails to resist motion (negative adhesion). With $\sigma_{sg} - \sigma_{sl} < 0$, viscous stresses displace the contact line without the mechanical resistance provided by wall adhesion. Under flow, this configuration cannot generate sufficient restoring force to stabilize the contact lines against viscous stresses, leading to breakup at $Ca = 0.03$ despite having the highest contact angles at lower capillary numbers.

\subsubsection{Normalized Wetting Displacement}
\begin{figure}
    \centering
    \includegraphics[width=0.6\textwidth]{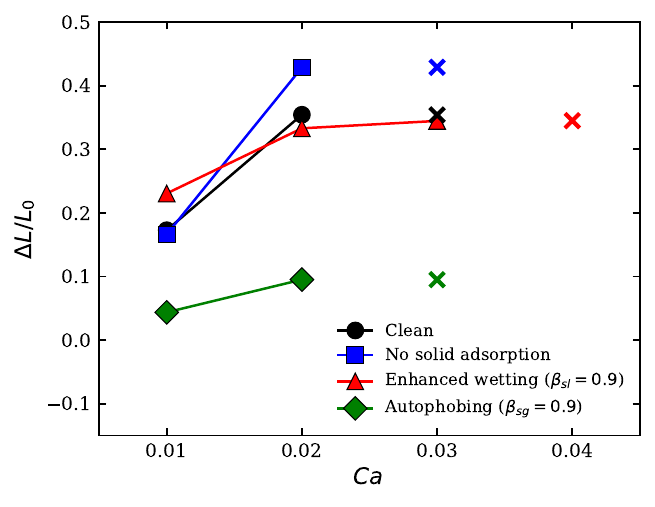}
    \caption{Normalized wetting displacement $\Delta L/L_0$ versus capillary number Ca for clean (black circles), no solid adsorption (blue squares), enhanced wetting with $\beta_{sl} = 0.9$ (red triangles), and autophobing with $\beta_{sg} = 0.9$ (green diamonds). The $\times$ markers indicate droplet breakup: clean, no solid adsorption, and autophobing fail at $\text{Ca} = 0.03$, while enhanced wetting survives until $\text{Ca} = 0.04$. }
    \label{fig:delta_wetting}
\end{figure}
 
To quantify droplet deformation under shear, we examine the normalized wetting displacement 
$\Delta L/L_0$, where $\Delta L = L_{\text{final}} - L_{\text{initial}}$. Figure \ref{fig:delta_wetting} presents this metric for $\text{Ca} = 0.01$ to $0.04$.

The results appear counterintuitive: autophobing exhibits the smallest displacement 
($\Delta L/L_0 \approx 0.04$ at $\text{Ca} = 0.01$) while enhanced wetting shows the largest ($\Delta L/L_0 \approx 0.23$). 
This ordering is misleading-the autophobing droplet displaces less because its higher 
equilibrium contact angle creates a smaller footprint exposed to shear, not because of 
greater mechanical resistance at the contact line.

Clean and no-solid-adsorption cases show nearly identical displacements at $\text{Ca} = 0.01$ 
($\Delta L/L_0 \approx 0.17$), but diverge at higher Ca: at $0.02$, no-solid-adsorption shows substantially 
larger displacement ($0.43$ vs $0.35$) due to reduced interfacial tension ($\sigma_{lg} \approx 0.9\sigma_0$). Lower 
interfacial tension reduces capillary pressure resisting deformation, allowing greater droplet 
deformation under identical viscous stress. However, unchanged solid--fluid interfacial tensions 
leave the adhesion force identical to clean-explaining simultaneous failure at $\text{Ca} = 0.03$.

This apparent paradox-larger displacement yet greater stability-is resolved by recognizing 
that equilibrium contact angle and adhesion force ($\sigma_{sg} - \sigma_{sl}$) are complementary. From 
Young's equation, $\sigma_{lg} \cos \theta_{eq} = \sigma_{sg} - \sigma_{sl}$: enhanced wetting reduces $\sigma_{sl}$ (increasing 
adhesion force, decreasing contact angle), while autophobing reduces $\sigma_{sg}$ (decreasing 
adhesion force, increasing contact angle). Smaller contact angles in enhanced wetting 
allow larger deformations before restoring force is overcome, while larger adhesion force 
provides greater detachment resistance. Thus, enhanced wetting droplets sustain larger 
displacements while remaining more stable.

\section{Conclusions}
\label{sec:con}
In this work, we have developed a thermodynamically consistent phase-field model for soluble surfactants in two-phase flows that incorporates surfactant adsorption on solid surfaces in addition to interfacial adsorption. The model is derived via variational principles consistent with the second law of thermodynamics, yielding modified free energies and kinetic boundary conditions that capture surfactant transport, adsorption dynamics, and wetting behaviour at moving contact lines.

A central contribution of this work is the identification of solid surface adsorption as the missing mechanism in prior numerical models of surfactant-laden contact line dynamics. Existing phase-field formulations that consider only interfacial adsorption predict that hydrophilic substrates become more hydrophilic and hydrophobic substrates more hydrophobic upon surfactant addition---a rescaling of Young's law that contradicts experimental observations. By incorporating surfactant adsorption at solid-liquid and solid-gas interfaces via the Langmuir--Szyszkowski equation of state, our model predicts a consistent shift toward increased hydrophilicity across all contact angles, consistent with experimental trends and qualitatively validated against contact angle measurements on PDMS surfaces with PEO and PEG solutions.

We have validated the model against analytical benchmarks for clean droplet spreading and derived scaling laws for the solid adsorption layer thickness in two distinct regimes: $\delta \propto \sqrt{\lambda_s}$ when wall accumulation is weak ($\Delta\psi \ll \psi_b$), and $\delta \propto \beta$ when wall affinity is strong ($\Delta\psi \gg \psi_b$). The concentration excess follows $\Delta\psi \propto \beta^2$ across all regimes. Numerical simulations confirm these scalings with excellent agreement.

The model also captures autophobing-a counterintuitive phenomenon wherein droplets exhibit increased contact angles upon surfactant addition. By reversing the asymmetry of solid adsorption strengths ($\beta_{sg} > \beta_{sl}$), we reproduce experimentally observed contact angle increases of approximately $15^\circ$, qualitatively matching measurements from \citet{bera2021antisurfactant}. The visualisation of asymmetric adsorption layers provides physical insight into the autophobing mechanism and demonstrates the generality of the solid adsorption framework.

Finally, we have extended the analysis to surfactant-laden droplets in Couette flow, illustrating how solid adsorption affects stability under shear. Reducing the fluid–fluid interfacial tension alone—through interfacial adsorption without solid adsorption—provides no stability benefit and can even increase deformation due to reduced capillary resistance. Only when surfactants adsorb preferentially on the solid–liquid interface (inside the droplet) does stability improve: enhanced wetting extends the critical capillary number by 20–30\%. In contrast, preferential adsorption on the solid–ambient fluid interface (autophobing) reduces stability despite producing higher contact angles.

These results establish that solid surface adsorption provides the missing mechanism required for predictive modelling of surfactant-laden contact line dynamics in the two-dimensional configurations studied here. The underlying thermodynamic argument — that wall adsorption modifies $\sigma_{sg}$ and $\sigma_{sl}$, thereby shifting Young's equation — does not rely on geometric dimensionality and is expected to carry over to three-dimensional systems. The framework therefore offers a foundation for investigating more complex configurations, including heterogeneous substrates, contact angle hysteresis effects, and three-dimensional geometries relevant to industrial coating, enhanced oil recovery, and biomedical applications. Furthermore, contact angle hysteresis can be incorporated within the same thermodynamically consistent framework, and this extension will be reported in a forthcoming study. The present simulations assume matched density and viscosity between the two fluid phases; extending the framework to systems with large property contrasts such as water--air is a natural next step.



\backsection[Acknowledgements]{The simulations were performed with computation time provided by the National Academic Infrastructure for Supercomputing in Sweden (NAISS) and the Swedish National Infrastructure for Computing (SNIC). We thank Dr. Armin Shahmardi for all the knowledge transfer.}

\backsection[Funding]{This work is supported by the funding provided by the European Research Council grant no. ‘2019-StG-852529, MUCUS’ and the Swedish Research Council through grant No 2021-04820.}

\backsection[Declaration of interests]{The authors report no conflict of interest.}


\backsection[Author ORCIDs]{

P. K. Kannan, https://orcid.org/0009-0004-1442-5082

K. T. Iqbal, https://orcid.org/0000-0003-1429-1008

D. Díaz, https://orcid.org/0000-0002-8904-6309

S. Mirjalili, https://orcid.org/0000-0003-4293-2431

G. Amberg, https://orcid.org/0000-0003-3336-1462

S. Bagheri, https://orcid.org/0000-0002-8209-1449

O. Tammisola, https://orcid.org/0000-0003-4317-1726}


\appendix

\section{Experimental Results}
\label{exp-res}

\begin{figure}
    \centering
    \includegraphics[width=0.8\textwidth]{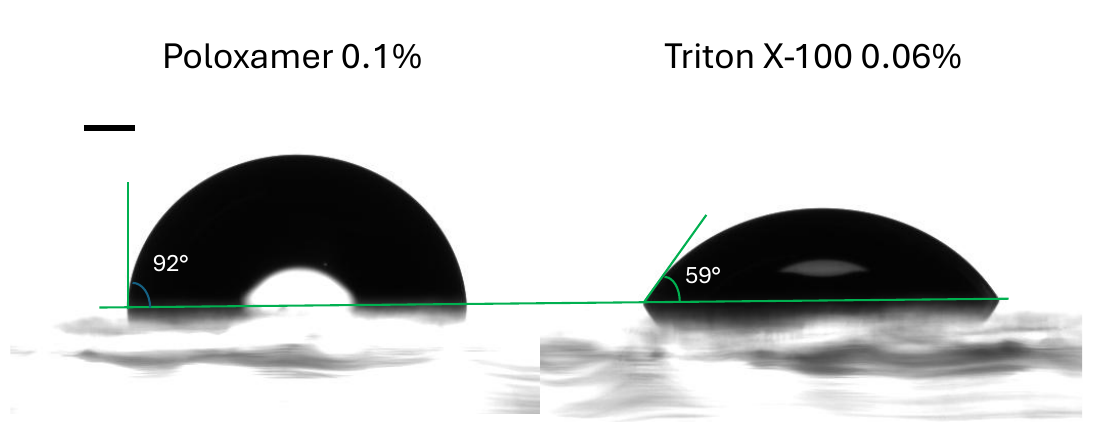}
    \caption{Decrease in equlibirium contact angle from $98^\circ$ for water (Fig \ref{fig:exp-st}) to $92^\circ$ for 0.1\% Poloxamer solution and $59^\circ$ for 0.06\% Triton X-100 solution on a PDMS substrate.}
    \label{fig:exp-app}
\end{figure}

In this section, we shall present the experiments with nonionic surfactants: Poloxamer (0.1 wt\%) and Triton X‑100 (0.06 wt\%) in water. Both surfactants produced a clear decrease in contact angle on the PDMS brush surface (8° for Poloxamer and 40° for Triton X‑100) as shown in Figure \ref{fig:exp-app}.

\section{Variational Derivation of CHNS Equations}
\label{app:chns_derivation}

\subsection{Variational Derivative and Equation of Motion}

We compute how $F$ changes under a small perturbation $\delta\phi$:
\begin{equation}
  \delta F = \lambda \int_V \left[ F_0'(\phi)\,\delta\phi
    + \nabla\phi \cdot \nabla(\delta\phi) + F_1'\delta\phi + F_\mathrm{Ex}'\delta\phi\right] dV
\end{equation}
Applying the divergence theorem to the gradient term:
\begin{equation}
  \delta F = \lambda \int_V \bigl[F_0'(\phi) - \nabla^2\phi + F_1' + F_\mathrm{Ex}'\bigr]\,\delta\phi\;dV
           + \lambda \oint_{\partial V} (\mathbf{n}\cdot\nabla\phi)\,\delta\phi\;dS
\end{equation}
The chemical potential is defined as the variational derivative:
\begin{equation}
  G_\phi \equiv \frac{\delta F}{\delta \phi}
  = \lambda\!\left[F_0'(\phi) - \nabla^2\phi + F_1' + F_\mathrm{Ex}'\right]
\end{equation}
This is the Fr\'echet derivative of $F$ with respect to $\phi$, 
i.e.\ $G_\phi = \delta F / \delta\phi$.
Since $\phi$ is conserved, we write $\partial\phi/\partial t + \nabla\cdot\mathbf{J}_\phi = 0$.
The rate of change of $F$ is:
\begin{equation}
  \frac{dF}{dt}
  = -\int_V \nabla G_\phi \cdot \mathbf{J}_\phi\;dV
\end{equation}
where boundary terms vanish by the no-flux condition. To guarantee 
$dF/dt \leq 0$, we choose:
\begin{equation}
  \mathbf{J}_\phi = M_\phi\,\nabla G_\phi
\end{equation}
with $M_\phi > 0$, giving $dF/dt = -M_\phi\int_V |\nabla G_\phi|^2\,dV \leq 0$.
Substituting into the conservation law yields the Cahn--Hilliard equation:
\begin{equation}
  \frac{\partial\phi}{\partial t} + \mathbf{u}\cdot\nabla\phi
  = M_\phi\,\nabla^2 G_\phi
\end{equation}
The analogous procedure applied to $\psi$ yields equation~(2.16).

\subsection{Derivatives of Free Energy Terms}
\label{app:fe_derivs}

From equations \eqref{F1} and \eqref{Fex}:
\begin{equation}
F_1 = -\frac{\lambda}{4\epsilon^2}\psi(1-\phi^2)^2 \quad \Rightarrow \quad F_1' = -\frac{\lambda}{4\epsilon^2}\psi \cdot \frac{\partial}{\partial\phi}\left[(1-\phi^2)^2\right] = \frac{\lambda\psi\phi(1-\phi^2)}{\epsilon^2},
\end{equation}
\begin{equation}
F_\text{Ex} = \frac{\lambda}{2\text{Ex}\cdot\epsilon^2}\psi\phi^2 \quad \Rightarrow \quad F_\text{Ex}' = \frac{\lambda\psi}{2\text{Ex}\cdot\epsilon^2} \cdot 2\phi = \frac{\lambda\psi\phi}{\text{Ex}\cdot\epsilon^2}.
\end{equation}
These combine into the final form of $G_\phi$ given in equation \eqref{eq:mu_phi}.

\subsection{Analogous Derivation for Surfactant}
\label{app:psi_deriv}

Applying the same procedure to $\psi$ (preserving the bulk free energy terms from equation \eqref{eq:Fpsi}):
\begin{align}
\delta \mathcal{F} &= \lambda_s \int_V (\bnabla\psi \cdot \bnabla\delta\psi) \drm V \\
&\quad + \frac{\lambda}{\epsilon^2}\int_V \left[\text{Pi}\ln\left(\frac{\psi}{1-\psi}\right)\delta\psi + \frac{\phi^2}{2Ex}\delta\psi - \frac{(1-\phi^2)^2}{4}\delta\psi\right] \drm V.
\end{align}

After integration by parts on the gradient term:
\begin{align}
\delta \mathcal{F} &= -\lambda_s \int_V \bnabla^2\psi \, \delta\psi \drm V \\
&\quad + \frac{\lambda}{\epsilon^2}\int_V \left[\text{Pi}\ln\left(\frac{\psi}{1-\psi}\right) + \frac{\phi^2}{2Ex} - \frac{(1-\phi^2)^2}{4}\right]\delta\psi \drm V \\
&\quad + \text{boundary terms}.
\end{align}

With flux form $\partial\psi/\partial t + \bnabla \cdot \boldsymbol{J}_\psi = 0$ and thermodynamic constraint $\boldsymbol{J}_\psi = M_\psi \bnabla G_\psi$ and choosing a degenerate mobility, we obtain:
\begin{equation}
\Ddt{\psi} = \bnabla \cdot (M_\psi \bnabla G_\psi),
\label{eq:transport_psi_final}
\end{equation}
with $G_\psi$ given by equation \eqref{eq:mu_psi}.

\subsection{Coupling to Navier-Stokes}
\label{app:ns_coupling}

The surface tension contributions $G_\phi \bnabla\phi$ and $G_\psi \bnabla\psi$ in the Navier-Stokes momentum balance (equation \eqref{eq:ns_momentum}) arise from the variational derivative of interfacial energy with respect to velocity. This follows the standard procedure of \cite{Jacqmin2000} and is not elaborated further here.

\section{Switch function $g(\phi)$ with surfactants} \label{sec:switch}

Following \citet{Yue2020cah}, we consider a planar diffuse interface intersecting the wall at angle $\theta_S$ under static equilibrium. This approach allows us to determine the form of $g(\phi)$ that ensures thermodynamic consistency in the presence of surfactants.

The chemical potential $G_\phi$ must be spatially uniform at equilibrium. For a one-dimensional interface with coordinate $\xi$ normal to the interface:
\begin{equation}
G_\phi = \lambda\left[-\frac{d^2\phi}{d\xi^2} + f_0'(\phi) + \frac{\psi_b\phi}{\text{Ex}\epsilon^2} + \frac{\psi_b\phi(1-\phi^2)}{\epsilon^2}\right] = 0,
\end{equation}
where $\psi_b$ is the bulk surfactant concentration and the last two terms arise from the variational derivatives of $F_1$ and $F_\text{Ex}$ (equations 2.5--2.6). This condition ensures that the phase field is in local equilibrium with respect to variations in $\phi$ while accounting for surfactant effects.

Multiplying by $d\phi/d\xi$ and integrating from $\xi = -\infty$ to $\xi$, using boundary conditions $f_0 = 0$, $d\phi/d\xi = 0$, and $\phi = \pm 1$ at infinity, yields:
\begin{equation}
-\frac{1}{2}\left(\frac{d\phi}{d\xi}\right)^2 + f_0 + \frac{\psi_b}{2\text{Ex}\epsilon^2}(\phi^2-1) - \frac{\psi_b}{4\epsilon^2}(\phi^4-1) + \frac{\psi_b}{2\epsilon^2}(\phi^2-1) = 0.
\end{equation}

When a planar interface intersects the wall with angle $\theta_S$, we get

\begin{equation}    \label{eq:uncom-stress}
    \lambda \boldsymbol{n} \cdot \bnabla \phi = \lambda \cos \theta_S \frac{\drm \phi}{\drm \xi} \; ,
\end{equation}
whereas equilibrium at the contact line means zero wall potential $L$,
\begin{equation}    \label{eq:fw-dash}
L(\theta,\phi,\bnabla \phi)=\lambda \boldsymbol{n} \cdot \bnabla \phi + 
f'_w(\phi)= 0\; ,
\end{equation}
This brings us to the form of the surface energy per unit area derivative in the presence of surfactants to be

\begin{equation}    \label{eq:fdashw-surf}
    f'_w(\phi) = -\frac{3}{2\sqrt{2}}\sigma \cos\theta_S\sqrt{ \frac{(\phi^2-1)^2}{2}   + \left(\frac{\psi_b}{\text{Ex} } +\psi_b\right)(\phi^2-1)   - \frac{1}{2}\psi_b(\phi^4-1)}  \;.
\end{equation}

Therefore we get 
\begin{equation}    \label{eq:gdash}
    g'(\phi) =  \frac{3}{2\sqrt{2}}\sqrt{ \frac{(\phi^2-1)^2}{2}  + \left(\frac{\psi_b}{\text{Ex} } +\psi_b\right)(\phi^2-1)   - \frac{1}{2}\psi_b(\phi^4-1)} \;.
\end{equation}
To obtain $g(\phi)$ from $g'(\phi)$, we integrate numerically and approximate with a 7th-order polynomial satisfying the boundary conditions $g(\pm 1) = \pm 0.5$:
\begin{equation}
g(\phi) = \frac{3}{2\sqrt{2}}\sqrt{A} \left[\phi + \frac{B}{6A}\phi^3 + \frac{C}{10A}\phi^5 + \frac{D}{14A}\phi^7\right],
\end{equation}
where
\begin{subequations}
\begin{align}
A &= \frac{1}{2} + \frac{\psi_b}{2} - \frac{\psi_b}{E_x} - \psi_b,\\
B &= -1 + \frac{\psi_b}{E_x} + \psi_b,\\
C &= \frac{1}{2} - \frac{\psi_b}{2},\\
D &= 14A\left[\frac{\sqrt{2}}{3\sqrt{A}} - 1 - \frac{B}{6A} - \frac{C}{10A}\right].
\end{align}
\end{subequations}

When $\psi_b \to 0$ (no surfactant), this reduces to the clean case $g(\phi) = \phi(3-\phi^2)/4$. The surfactant-dependent terms modify the switch function to reflect changes in interfacial energetics induced by adsorption.

\section{Langmuir isotherm approximation for fluid-fluid surface tension}
\label{ap:flfl}
To validate the surfactant behavior at fluid-fluid interfaces, we perform spreading droplet simulations by keeping the clean case equilibirum angle constant but varying $\psi_b$, using the same model but with solid adsorption disabled ($\beta_{sl} = \beta_{sg} = 0$). In this configuration, the solid surface tensions remain unchanged, and Young's equation reduces to:
\begin{equation}
    \sigma_e \cos\theta_e = \sigma_0 \cos\theta_0
\end{equation}
where $\sigma_0$ and $\theta_0$ are the surface tension and contact angle of the clean interface. This allows us to extract the equilibrium surface tension $\sigma_e$ directly from the measured contact angle $\theta_e$.

\begin{figure}
    \centering
    \includegraphics[width=0.8\textwidth]{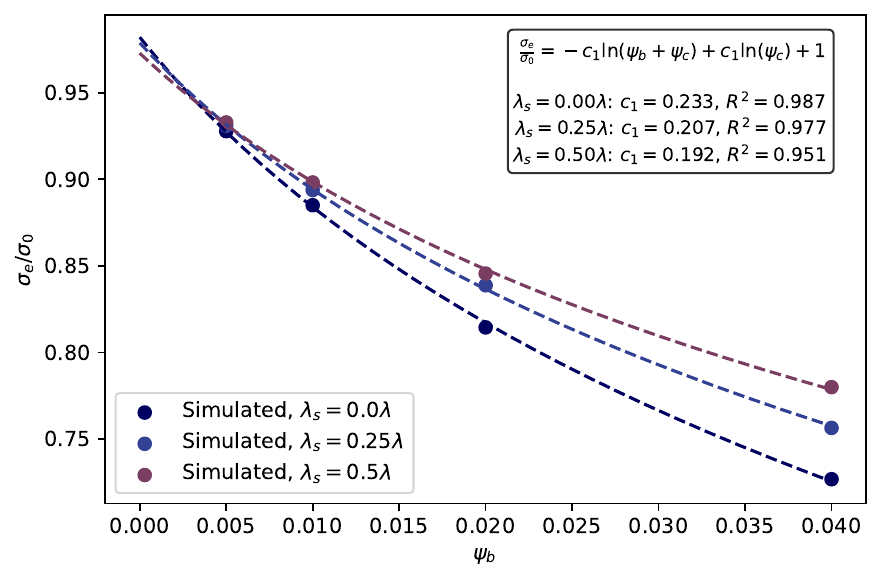}
    \caption{Relative surface tension $\sigma_e/\sigma_0$ as a function of bulk surfactant concentration $\psi_b$ for different values of the gradient penalty parameter $\lambda_s$. Symbols denote simulation results; dashed lines show fits to the Langmuir isotherm. The inset displays the fitted equation and corresponding coefficients $c_1$ and $R^2$ values for each $\lambda_s$.}
    \label{fig:isotherm_validation}
\end{figure}

Figure~\ref{fig:isotherm_validation} shows $\sigma_e/\sigma_0$ as a function of bulk surfactant concentration $\psi_b$ for different values of the gradient penalty parameter $\lambda_s$. The simulation results are fitted to the Langmuir isotherm:
\begin{equation}
    \frac{\sigma_e}{\sigma_0} = 1 + c_1 \ln\left(\frac{\psi_c}{\psi_b + \psi_c}\right)
\end{equation}

The high $R^2$ values (0.95-0.99) confirm that the inclusion of the square gradient term in our free energy formulation does not cause significant deviation from the Langmuir isotherm, consistent with \citet{Engblom2013}. As expected, $R^2$ decreases with increasing $\lambda_s$, reflecting the growing influence of the gradient penalty on interfacial surfactant distribution. Throughout this work, we use $\lambda_s = 0.5\lambda$, for which $R^2 = 0.951$ and $c_1 = 0.192$. We conclude that for this value of $\lambda_s$, the fluid-fluid interface follows the Langmuir isotherm to a good approximation.

\section{Experimental Methods}
\label{sec:exp-methods}

In this section, we shall discuss the method used to measure the static angles when surfactants are added to the system (Figure \ref{fig:exp-st}) to compare them qualitatively with numerical studies.
\subsection{Surface preparation}

The preparation of PDMS brushes are based on the methods reported by \cite{liu2021one}. Glass microscope slides (26~$\times$~76~mm$^{2}$, 1~mm thick) were Oxygen-plasma-treated for 5 min at 70 W (Diener Electronic Plasma Surface Technology: Femto BLS, Ebhausen, Germany). Afterwards, the substrates were immersed in 40 mL of water-saturated toluene (0.024 mM) containing 1.4~mL of Dichlorodimethylsilane (DCDMS, Sigma–Aldrich) for 30 min to get PDMS-brushed-coated surfaces.  After reaction, the samples were rinsed with Toluene.
\subsection{Surfactant solutions}
Polyethylene oxide (PEO 8~$\times$~10$^{6}$, Sigma Aldrich) and Polyethylene Glycol (PEG 8000, MV 7000-9000, Fisher Bioreagents) solutions were prepared in concentrations of 0.1~$\%$ and 0.3~$\%$ in water. The solutions were stirred by a magnetic stirrer for 30 min at 1100 rpm.
\subsection{Contact angle measurements}
Contact angles were measured by a Goniometer Drop Shape Analyzer (DSA25, Krüss). The static contact angle was measured by the sessile drop method, depositing an 8~$\mu$L drop of deionized water (18~M$\Omega$) on the surface on four different spots. A tangent fitting method provided by the ADVANCE software of the device was used to determine the static contact angle.

\bibliographystyle{jfm}
\bibliography{jfm}

\end{document}